\documentclass[a4paper,fleqn,usenatbib]{mnras}
\usepackage{latexsym}
\usepackage{amsmath}
\usepackage{amssymb}
\usepackage{fnpos}
\usepackage{hyperref}
\usepackage{tabularx}
\usepackage{xspace}
\usepackage{enumerate}
\usepackage{scalerel}
\usepackage{graphicx}
\usepackage{url}
\usepackage{caption}
\usepackage{gensymb}
\usepackage{longtable}
\usepackage{lscape}
\usepackage[normalem]{ulem} 
\usepackage[dvipsnames]{xcolor} 
\usepackage{wasysym} 
\definecolor{darkgreen}{rgb}{0.0,0.55,0.0}
\definecolor{darkblue}{rgb}{0.0,0.0,0.5}

\newcommand{\eal}[2]{\ifmmode{\mathrm{#1\,#2}}\else{#1\textsc{$\,$\lowercase{#2}}}\fi\xspace}
\newcommand{\feal}[2]{\ifmmode{\mathrm{#1\,#2}}\else{[#1\textsc{$\,$\lowercase{#2}}]}\fi\xspace}
\newcommand{\hfeal}[2]{\ifmmode{\mathrm{#1\,#2}}\else{#1\textsc{$\,$\lowercase{#2}}]}\fi\xspace}

\newcommand{\orcid}[1]{$^{\rm \href{https://orcid.org/#1}{\includegraphics[height=0.6em]{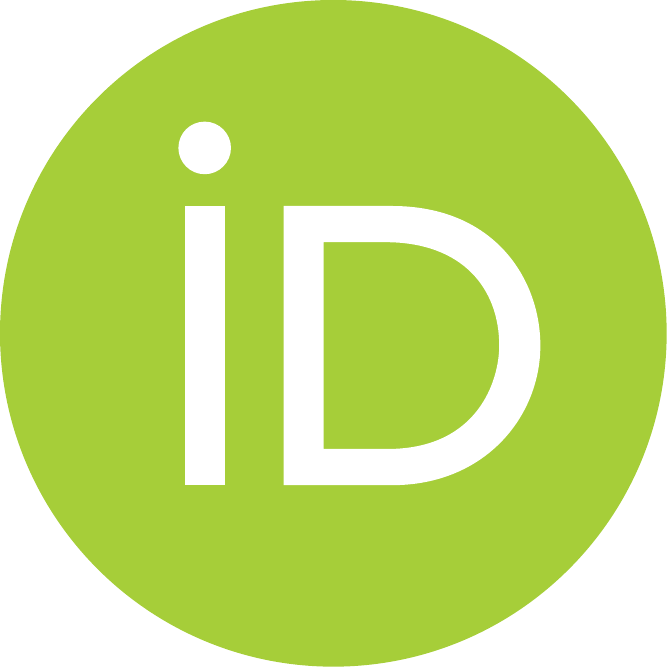}}}$}

\title[The slow nova Gaia22alz]{Catching a nova X-ray/UV flash in the visible? Early spectroscopy of the extremely slow Nova~Velorum~2022 (Gaia22alz)}
\author[Aydi et al.]{\parbox{\textwidth}{E.~Aydi\orcid{0000-0001-8525-3442}$^{1}$\thanks{Elias Aydi - NFHP Hubble Fellow; E-mail: aydielia@msu.edu}, L.~Chomiuk\orcid{0000-0002-8400-3705}$^{1}$, 
J.~Miko{\l}ajewska\orcid{0000-0003-3457-0020}$^{2}$, J.~Brink$^{3, 4}$, B.~D.~Metzger\orcid{0000-0002-4670-7509}$^{5, 6}$, J.~Strader$^{1}$, D.~A.~H.~Buckley$^{3, 4}$, E.~J.~Harvey\orcid{0000-0002-3014-3665}$^{7}$, T.~W.-S.~Holoien$^{8}$, L.~Izzo\orcid{0000-0001-9695-8472}$^{9}$, A.~Kawash$^{1}$, J.~D.~Linford\orcid{0000-0002-3873-5497}$^{10}$, P.~Molaro$^{11, 12}$, B.~Mollina$^{1}$, P.~Mr\'oz\orcid{0000-0001-7016-1692}$^{13}$, K.~Mukai$^{14, 15}$, M.~Orio$^{16, 17}$, T.~Panurach\orcid{0000-0001-8424-2848}$^{1}$, P.~Senchyna$^{8}$, B.~J.~Shappee$^{18}$, K.~J.~Shen\orcid{0000-0003-4631-1149}$^{19}$, J.~L.~Sokoloski$^{20}$, K.~V.~Sokolovsky\orcid{0000-0001-5991-6863}$^{21, 22}$, R.~Urquhart$^{1}$, and R.~E.~Williams$^{23, 24}$}
\vspace{0.4cm}\\
\parbox{\textwidth}{
$^{1}$Center for Data Intensive and Time Domain Astronomy, Department of Physics and Astronomy, Michigan State University, East Lansing, MI 48824, USA\\
$^{2}$Nicolaus Copernicus Astronomical Center, Polish Academy of Sciences, Bartycka 18, PL 00-716 Warsaw, Poland\\
$^{3}$South African Astronomical Observatory, P.O. Box 9, 7935 Observatory, South Afric\\
$^{4}$Department of Astronomy, University of Cape Town, Private Bag X3, Rondebosch 7701, South Africa\\
$^{5}$Department of Physics and Columbia Astrophysics Laboratory, Columbia University, New York, NY 10027, USA\\
$^{6}$Center for Computational Astrophysics, Flatiron Institute, 162 5th Ave, New York, NY 10010, USA\\
$^{7}$UK Astronomy Technology Centre, Royal Observatory, Blackford Hill, Edinburgh, EH9 3HJ, UK \\
$^{8}$Carnegie Observatories, 813 Santa Barbara Street, Pasadena, CA 91101, USA\\
$^{9}$DARK, Niels Bohr Institute, University of Copenhagen, Jagtvej 128, 2200 Copenhagen {\O}, Denmark \\
$^{10}$National Radio Astronomy Observatory, P.O. Box O, Socorro, NM 87801, USA\\
$^{11}$INAF-Osservatorio Astronomico di Trieste, Via G.B. Tiepolo 11, I-34143 Trieste, Italy\\
$^{12}$Institute of Fundamental Physics of the Universe, Via Beirut 2, Miramare, Trieste, Italy\\
$^{13}$Astronomical Observatory, University of Warsaw, Al. Ujazdowskie 4, 00-478 Warszawa, Poland\\
$^{14}$CRESST and X-ray Astrophysics Laboratory, NASA/GSFC, Greenbelt, MD 20771, USA\\
$^{15}$Department of Physics, University of Maryland, Baltimore County, 1000 Hilltop Circle, Baltimore, MD 21250, USA\\
$^{16}$INAF--Osservatorio di Padova, vicolo dell'Osservatorio 5, I-35122 Padova, Italy\\
$^{17}$Department of Astronomy, University of Wisconsin, 475 N.\ Charter St., Madison, WI 53704, USA \\
$^{18}$Institute for Astronomy, University of Hawai’i, 2680 Woodlawn Drive, Honolulu, HI 96822, USA\\
$^{19}$Department of Astronomy and Theoretical Astrophysics Center, University of California, Berkeley, CA 94720, USA\\
$^{20}$Columbia Astrophysics Laboratory and Department of Physics, Columbia University, New York, NY 10027, USA\\
$^{21}$Department of Astronomy, University of Illinois at Urbana-Champaign, 1002 W. Green Street, Urbana, IL 61801, USA\\
$^{22}$Sternberg Astronomical Institute, Moscow State University, Universitetskii pr. 13, 119992 Moscow, Russia\\
$^{23}$Department of Astronomy \& Astrophysics, University of California, Santa Cruz, 1156 High Street, Santa Cruz, CA 95064, USA\\
$^{24}$Space Telescope Science Institute, 3700 San Martin Drive, Baltimore, MD 21218, US\\
}}

\pubyear{2023}

\begin{document}
\label{firstpage}
\pagerange{\pageref{firstpage}--\pageref{lastpage}}
\maketitle

\begin{abstract}
We present early spectral observations of the very slow Galactic nova Gaia22alz, over its gradual rise to peak brightness that lasted 180 days. During the first 50 days, when the nova was only 3--4~magnitudes above its normal brightness, the spectra showed narrow (FWHM $\approx$ 400\,km\,s$^{-1}$) emission lines of H Balmer, \eal{He}{I}, \eal{He}{II}, and \eal{C}{IV}, but no P~Cygni absorption. A few weeks later, the high-excitation \eal{He}{II} and \eal{C}{IV} lines disappeared, and P~Cygni profiles of Balmer, \eal{He}{I}, and eventually \eal{Fe}{II} lines emerged, yielding a spectrum typical of classical novae before peak. We propose that the early spectra of Gaia22alz are produced in the white dwarf's envelope or accretion disk, reprocessing X-ray and ultraviolet emission from the white dwarf after a dramatic increase in the rate of thermonuclear reactions, during a phase known as the ``early X-ray/UV flash''. If true, this would be one of the rare times that the optical signature of the early X-ray/UV flash has been detected. While this phase might last only a few hours in other novae and thus be easily missed,  it was possible to detect in Gaia22alz due to its very slow and gradual rise and thanks to the efficiency of new all-sky surveys in detecting transients on their rise. We also consider alternative scenarios that could explain the early spectral features of Gaia22alz and its unusually slow rise.
\end{abstract}

\begin{keywords}
stars: novae, cataclysmic variables --- white dwarfs.
\end{keywords}

\section{Introduction}
Classical Novae are thermonuclear runaway events taking place on the surface of accreting white dwarfs in interacting binary star systems (for reviews see, e.g., \citealt{Bode_etal_2008,Della_Valle_Izzo_2020,Chomiuk_etal_2020}). 
The eruption leads to an increase in the visible brightness of the system by 8 to 15 magnitudes over a few hours to days. While most classical novae rapidly ascend to peak brightness, some exhibit a prolonged brightening lasting for several weeks
(e.g., \citealt{Chochol_etal_2003,  Strope_etal_2010,Tanaka_etal_2011,Tanaka_etal_2011_159,Aydi_etal_2019_I,Aydi_etal_2020a}). 

During the rise to peak, the spectra of novae are typically dominated by P Cygni profiles of Balmer, \eal{Fe}{II}, \eal{He}{I}, \eal{O}{I}, and \eal{Na}{I}, characterized by velocities ranging between a few hundreds to $\approx$
1000\,km\,s$^{-1}$. After peak, faster emission components emerge with velocities of a few
1000\,km\,s$^{-1}$ \citep{McLaughlin_1944,McLaughlin_1947,Payne-Gaposchkin_1957,Aydi_etal_2020b}. It is challenging to observe novae before they reach optical peak, mainly due to their typically quick rise and their transient nature. However, slowly rising novae give us a chance to monitor them during the days -- weeks leading to their visible peak. Nevertheless, even the slowest novae are  discovered when they are already several magnitudes ($>$5\,mag) above quiescent brightness (e.g., \citealt{Strope_etal_2010,Hounsell_etal_2010,Hounsell_etal_2016}), with the exception of just a few (e.g., HR Del; \citealt{Friedjung_1992} and T Pyx; \citealt{Arai_etal_2015}). Therefore, the early times of the eruption are often missed, and spectral observations during these early phases are scarce. 

Gaia22alz (AT2022bpq; Nova Velorum 2022) was reported by Gaia science alerts\footnote{\url{http://gsaweb.ast.cam.ac.uk/alerts}} on 2022 Feb 04.25 UT as an optical transient at $G$ = 16.87 mag, towards the constellation Vela \citep{2022TNSTR.313....1H}. 
The transient is located at equatorial coordinates of $(\alpha, \delta)$ = (09$^{\mathrm{h}}$29$^{\mathrm{m}}$53$^{\mathrm{s}}$\!\!\!.\,06, --56$^{\circ}$17$'$25$''$\!\!\!.\,94) and Galactic coordinates of ($l, b$) = (277$^{\circ}$\!\!\!.\,563, --3$^{\circ}$\!\!\!.\,692). There is a Gaia EDR3 \citep{2021A&A...649A...1G} source (ID 5307057922488654848) matching the transient position with an average $G$ magnitude of 18.3. Archival photometry from the All-Sky Automated Survey for SNe (ASAS-SN; \citealt{Shappee_etal_2014,Kochanek_etal_2017}), showed that the eruption started as early as
%
%
2022 Jan 25.02~UT (HJD 2459604.522; \citealt{ATel_15355}); we will consider this date as $t_0$ in the remainder of the paper. In archival data from the Dark Energy Camera Plane Survey (DE-CaPS; \citealt{Schlafly_etal_2018}) the source has median pre-eruption magnitude of
$g \simeq 19.0$.

\citet{ATel_15270} reported spectroscopic observations taken around 45 days after $t_0$ (still during the early rise to peak), showing narrow emission features of \eal{H}{I} Balmer, \eal{He}{I}, \eal{He}{II}, and \eal{C}{IV}. Based on these features, and the then-low-amplitude outburst, \citet{ATel_15270} suggested that the transient might be 
an unusual disk instability (dwarf nova) event. Later observations, 
taken as the transient climbed to peak, were more consistent with a classical nova eruption, showing typical P Cygni profiles of Balmer, \eal{He}{I}, \eal{Fe}{II}, and \eal{O}{I} \citep{ATel_15355,ATel_15395}. 
This suggests that the early spectra 
of Gaia22alz are unique and, covering a phase of a nova eruption that is usually missed.

Here we present these early observations of nova Gaia22alz, focusing on its unusual spectral evolution. In Section~\ref{Obs} we present the observations and data reduction. In Section~\ref{Res} we show our results, while in Section~\ref{Disc} we offer discussion about the nature of the eruption and the origin of the early spectral features. Our conclusions are presented in Section~\ref{sec_conc}.

\begin{figure}
\begin{center}
\includegraphics[width=\columnwidth]{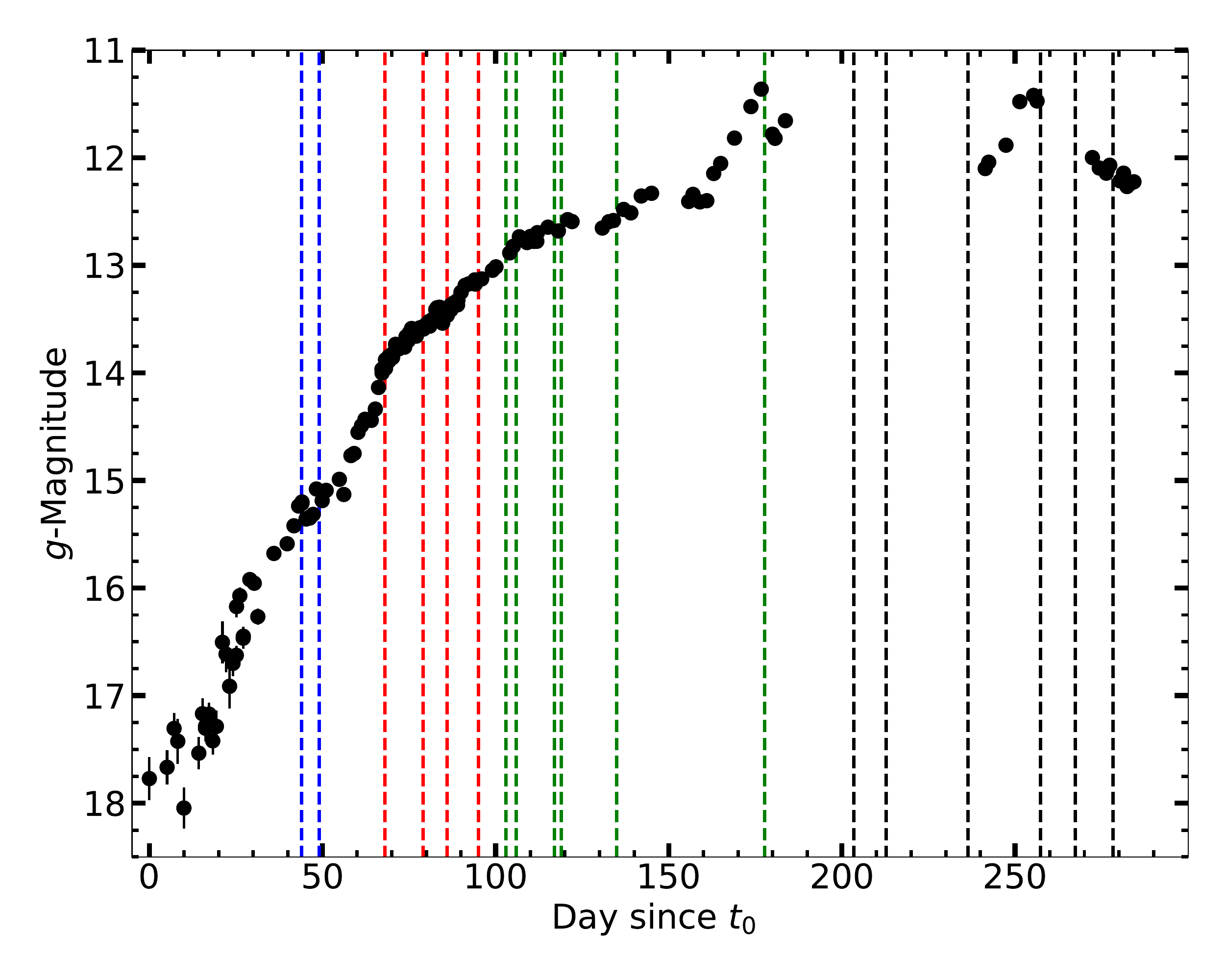}
\caption{The ASAS-SN optical $g$-band light curve of nova Gaia22alz during the first 300 days of the eruption, focusing on the rise to peak brightness. The dashed lines represents times of spectroscopic observations; blue, red, green, and black lines, represent different stages in the spectral evolution, namely, the early-rise, mid-rise, late-rise, and post-optical-peak, respectively (See text for more details).}
\label{Fig:LC}
\end{center}
\end{figure}

\section{Observations and data reduction}
\label{Obs}

We make use of publicly available photometry from the ASAS-SN survey. These data are taken as part of the all-sky patrol in the $g$-band. The data can be found in the supplementary online material.

We obtained optical spectroscopic observations of Gaia22alz between days 45 and 278 using a
variety of telescopes and instruments. A log of the spectral observations is presented in Table~\ref{table:spec_log}. Below we present details of these observations.

We obtained spectra using the High Resolution Spectrograph (HRS; \citealt{Barnes_etal_2008,Bramall_etal_2010,Bramall_etal_2012,Crause_etal_2014}) and the Robert Stobie Spectrograph (RSS; \citealt{Burgh_etal_2003}; \citealt{Kobulnicky_etal_2003}) mounted on the Southern African Large Telescope (SALT; \citealt{Buckley_etal_2006,Odonoghue_etal_2006}) in Sutherland, South Africa. We used HRS in the low resolution LR mode, yielding a resolving power $R = \lambda/\Delta\lambda  \approx$ 14,000 over the range 4000\,--\,9000\,\AA{}. The primary reduction of the HRS spectroscopy was conducted using the SALT science pipeline \citep{Crawford_etal_2010}, which includes over-scan correction, bias subtraction, and gain correction. The rest of the reduction was done using the MIDAS FEROS \citep{Stahl_etal_1999} and $echelle$ \citep{Ballester_1992} packages. The reduction procedure is described by \citet{kniazev_etal_2016}. RSS was used in long-slit mode with the 1.5 arcsec slit and the PG900 grating, resulting in a resolving power $R \approx 1500$. The spectra were first reduced using the  PySALT pipeline \citep{Crawford_etal_2010}, which involves bias subtraction, cross-talk correction, scattered light removal, bad pixel masking, and flat-fielding. The wavelength calibration, background subtraction, and spectral extraction were done using the Image Reduction and Analysis Facility (IRAF; \citealt{Tody_1986}).

We also carried out low- and medium-resolution optical spectroscopy using the Goodman spectrograph \citep{Clemens_etal_2004} on the 4.1\,m Southern Astrophysical Research (SOAR) telescope located on Cerro Pach\'on, Chile. The observations were carried out in two setups: one setup using the 400~l\,mm$^{-1}$ grating and a 1.2\arcsec\ slit, yielding a resolving power $R \approx$ 1100 over the wavelength range 3820--7850\,\AA{}. Another setup utilized 
a 2100~l\,mm$^{-1}$ grating and a 1.2\arcsec\ slit, yielding a resolving power $R \approx$ 4400 in a 570\,\AA\ wide region centered on H$\beta$. 
The spectra were reduced and optimally extracted using the \textsc{apall} package in IRAF.

We also used the  Inamori-Magellan Areal Camera \& Spectrograph (IMACS) on the Magellan-Baade 6.5\,m telescope at Las Campanas Observatory.
On the night of 2022-07-21 we obtained a spectrum in longslit mode with the 600 l/mm grating and the f/4 camera through a 0.9\arcsec\ slit, yielding a spectral resolution of $R \approx 1500$ over 3800--6600~\AA{}. The spectra were reduced and optimally extracted using the \textsc{apall} package in IRAF.

\begin{figure*}
\begin{center}
  \includegraphics[width=\textwidth]{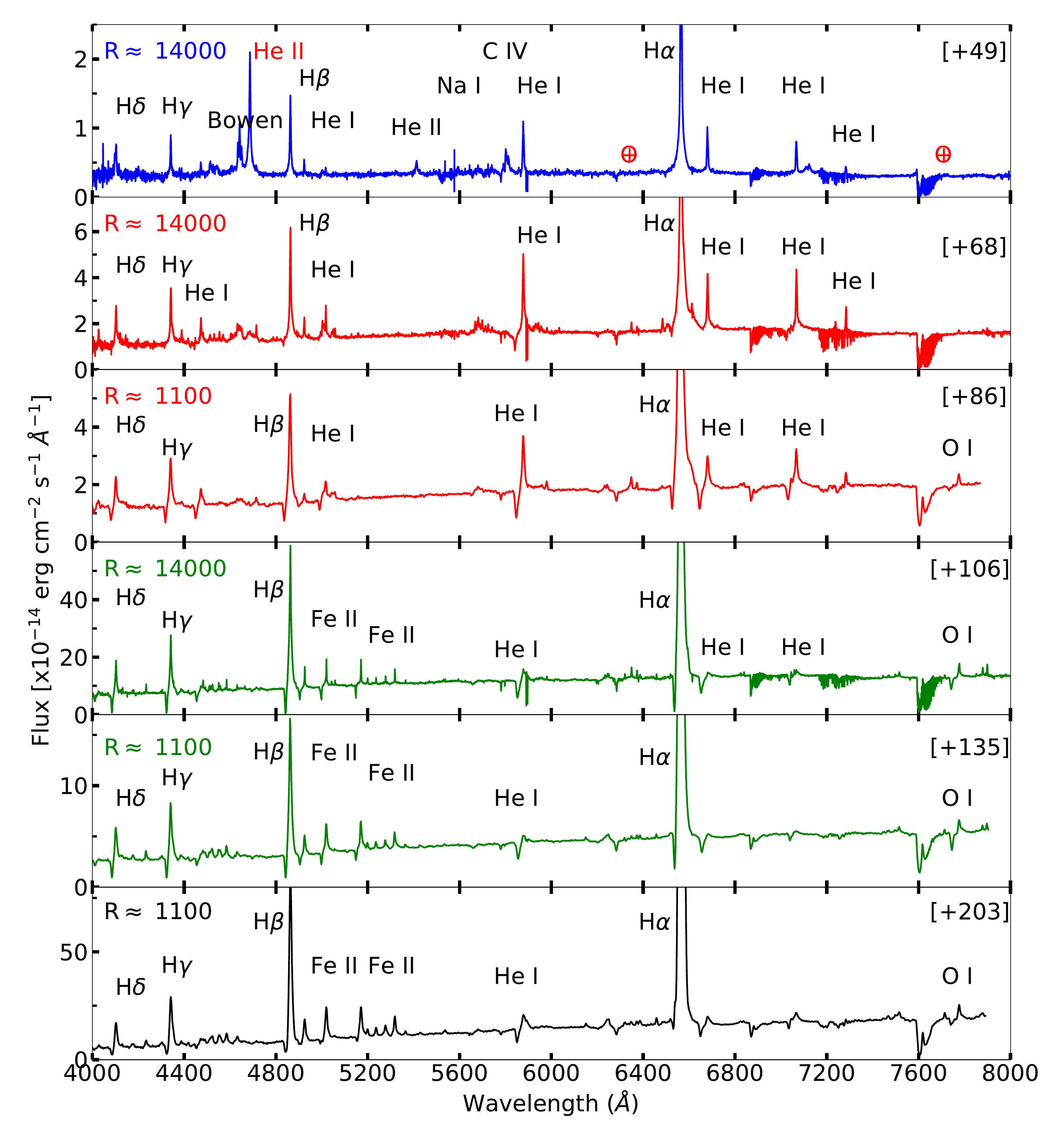}
\caption{The spectral evolution of nova Gaia22alz representing the different spectral stages:  early-rise (blue), mid-rise (red), late-rise (green), and post-optical-peak (black). Numbers between brackets are days after $t_0$. The resolving power $R$ of each spectrum is added on the left-hand side of each panel.}
\label{Fig:main_spec}
\end{center}
\end{figure*}

\section{Results}
\label{Res}

\subsection{Photometric evolution}
In Figure~\ref{Fig:LC} we present photometry of Gaia22alz from the ASAS-SN survey, focusing on the first 300 days of the eruption --- as the focus of this work is on the early spectral evolution of the nova. A complete light curve is presented in Figure~\ref{Fig:LC_comp}, showing the ASAS-SN photometry up until the time of writing of this paper. Solar conjunction prevented ASAS-SN from obtaining photometric observations between days $\sim$185--240 after $t_0$. Due to this gap in the monitoring, it is not possible to definitively constrain the date of optical peak in the light curve, nor its magnitude. However, it is possible that the peak was reached around day 177 at $g \gtrsim$ 11.2 mag, implying an eruption amplitude of $\gtrsim$ 7.8 magnitudes. 
This is at the lower end of eruption amplitudes distribution found in novae \citep[$\sim 8$--15\,mag;][]{1990ApJ...356..609V,Warner_2008,2021ApJ...910..120K}.

The rise of nova Gaia22alz is slow and gradual, taking more than 180 days to reach peak brightness. 
While other novae have taken a similarly long time to reach peak, their rises are not as gradual as observed in Gaia22alz; 
these other eruptions typically start with a rapid rise in brightness by 5 to 6 magnitudes, then at some point slow down. It takes a typical slow nova several weeks or months to climb the remaining 3 to 4 magnitudes
to its peak (e.g., V723~Cas: \citealt{Munari_etal_1996,Hachisu_Kato_2004,Goranskij_etal_2007}, V1548~Aql: \citealt{Kato_Takamizawa_2001}, V1280~Sco: \citealt{Hounsell_etal_2016}, and the recent example of V1405~Cas). 

In contrast, Gaia22alz was detected less than a magnitude above its quiescence level, showing a slow-paced rise to optical peak. While for many slow novae the durations of their initial fast rise are well constrained, it cannot be excluded that some historical slow novae also could have shown behavior similar to that of Gaia22alz (e.g., HR Del; \citealt{Strope_etal_2010}).
As deep high-cadence all-sky photometric and transient surveys like ASAS-SN and Gaia have become operational only recently, the future is bright for discovering more slow novae at a stage when they are just above their quiescent brightness.

Some time around or during solar conjunction (days $\sim180-240$), the optical light curve of Gaia22alz peaked, and this peak was accompanied by at least one additional maximum/flare 
(e.g., the secondary maximum on day 255; see Figure~\ref{Fig:LC}).
Many novae have been observed to show multiple maxima/flares after peaking (e.g., \citealt{Pejcha_2009,Strope_etal_2010,Tanaka_etal_2011_159,Tanaka_etal_2011,Walter_2016,Shore_etal_2018,Aydi_etal_2019_I,Aydi_etal_2020a}), particularly amongst those with slowly evolving light curves. Novae are divided into different speed classes based on the rate of decline of the light curve by two magnitudes from peak, known as the $t_2$ timescale. We consider $t_2$ as the last time the light curve dropped 2 magnitudes below peak brightness \citep{Strope_etal_2010}. Assuming that we captured the light curve peak on day 177, we derive $t_2 \approx$ 212\,days (Figure~\ref{Fig:LC_comp}), meaning that Gaia22alz belongs to the class of very slow novae \citep{Payne-Gaposchkin_1957}. However, at the time of writing of this paper, the nova was still showing flares and it might still rise above 2 magnitudes below peak (see Figure~\ref{Fig:LC_comp}).

Photometric measurements taken from the American Association of Variable Star Observers (AAVSO; \citealt{Kafka_2020}) on day 177 (near optical peak) place the nova at $V=10.8$ mag. Based on this value and assuming a reddening $A_V$ = 3.5\,mag and a distance $\geq$ 4.5\,kpc towards Gaia22alz (see Section~\ref{sec_dist}), we derive a peak absolute magnitude $M_V < -5.9$. which is within the typical range of peak absolute magnitudes of classical novae:
$M_V = -5$ to $-11$\,mag \citep{2017ApJ...834..196S,2022MNRAS.517.6150S}.

\subsection{Spectral evolution}
In Figure~\ref{Fig:main_spec} we present the overall spectral evolution of Gaia22alz throughout the eruption. Based on the spectral features, we divide the spectral evolution into four chronological stages. These stages are highlighted with different colors in Figure~\ref{Fig:LC} and \ref{Fig:main_spec}: (1) early-rise (highlighted in blue); (2) mid-rise (highlighted in red); (3) late-rise (highlighted in green); (4) post-optical-peak (highlighted in black). A complete spectral evolution is presented in Figures~\ref{Fig:spec_evolution_1} to~\ref{Fig:spec_evolution_3}.

\begin{figure*}
\begin{center}
  \includegraphics[width=0.8\textwidth]{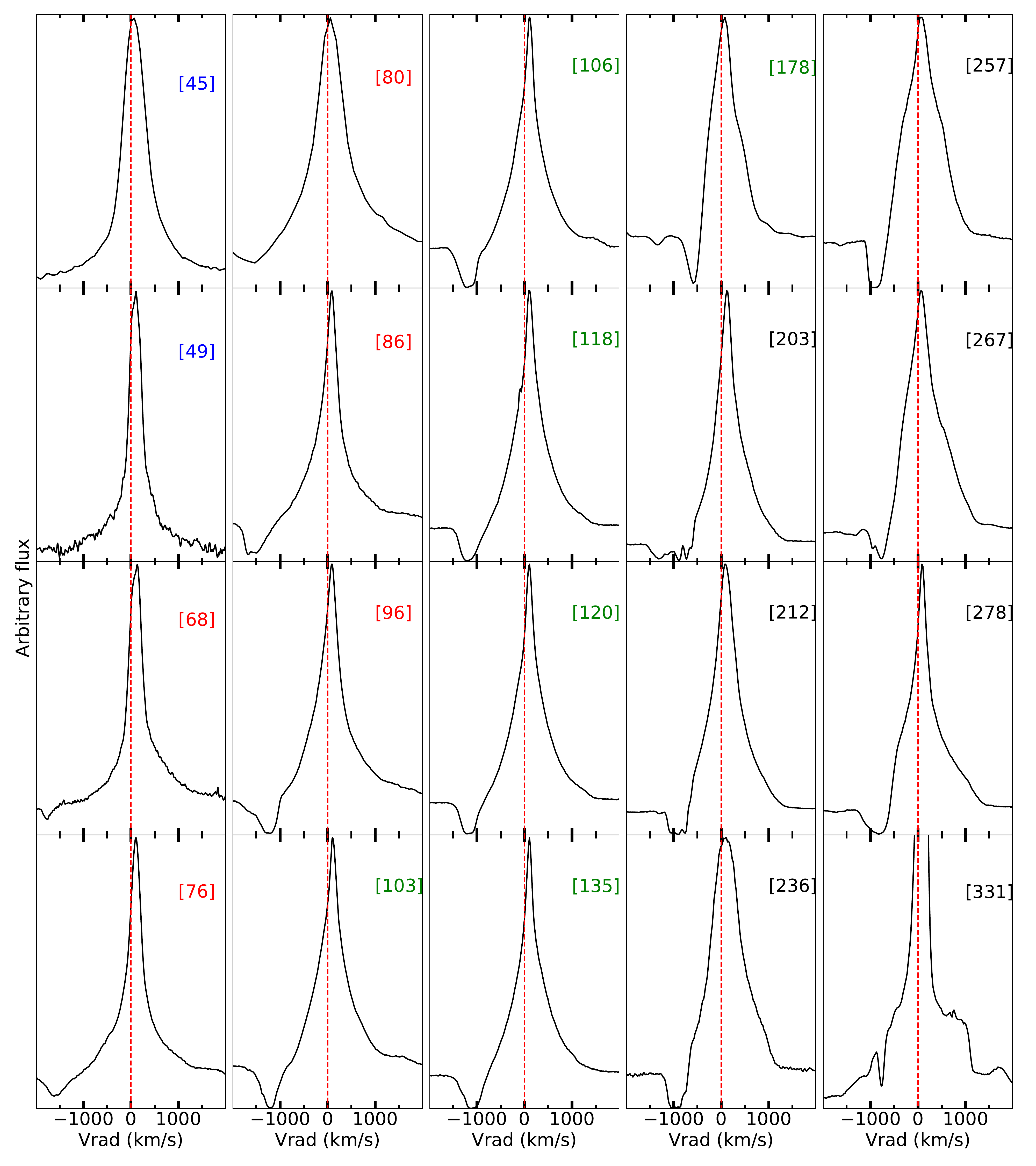}
\caption{The line profile evolution of H$\beta$ during the first 300 days of the eruption. The red dashed line represent rest velocity ($v_{\mathrm{rad}} = 0$\,km\,s$^{-1}$). Note that the spectra on days 45 and 80 have lower resolution ($R \approx 1000$), compared to the other spectral epochs ($R \approx 4400$ or 14,000).} 
\label{Fig:Hbeta}
\end{center}
\end{figure*}

The first two spectra obtained on days 45 and 49 during the early-rise are not typical of novae. 
The spectra are dominated by relatively narrow (FWHM $\approx$ 400\,km\,s$^{-1}$) emission lines of H Balmer and \eal{He}{I}, along with \eal{He}{II} 4686 and 5412\,\AA{}, \eal{C}{IV}, and the Bowen blend. The Balmer lines are characterized by broad wings extending to around 2500\,km\,s$^{-1}$ (see Figure~\ref{Fig:Hbeta}). We measure a flux ratio $F_{He II}/F_{H\beta} \approx 2$ from both spectra, highlighting the strength of the \eal{He}{II} emission line at 4686\,\AA{}, which indicates the presence of a hot emitting source ($T \approx$ a few times 10$^{5}$\,K; e.g., \citealt{Morisset_Pequignot_1996,Bonning_etal_2013}). The narrow lines and the absence of P Cygni profiles are not consistent with optically-thick ejecta typically observed in novae before peak \citep{Aydi_etal_2020b}. The spectra at this stage are reminiscent of the early spectra of the 2011 eruption of nova T Pyx which was obtained by \citet{Arai_etal_2015} only 4.4 hours after the eruption discovery. The early spectra of T Pyx were also dominated by Balmer and high-excitation lines of \eal{He}{II}, \eal{N}{III}, and \eal{C}{IV} with no P Cygni profiles. However, the line were characterized by high velocities, typical of classical novae (FWHM $\approx$ 1100--1500\,km\,s$^{-1}$). \citet{Arai_etal_2015} suggested that at this early stages, the spectrum implies that the envelope of the WD would be optically thin and the optically thick wind/outflow is still weak. 
\\

\begin{figure}
\begin{center}
  \includegraphics[width=\columnwidth]{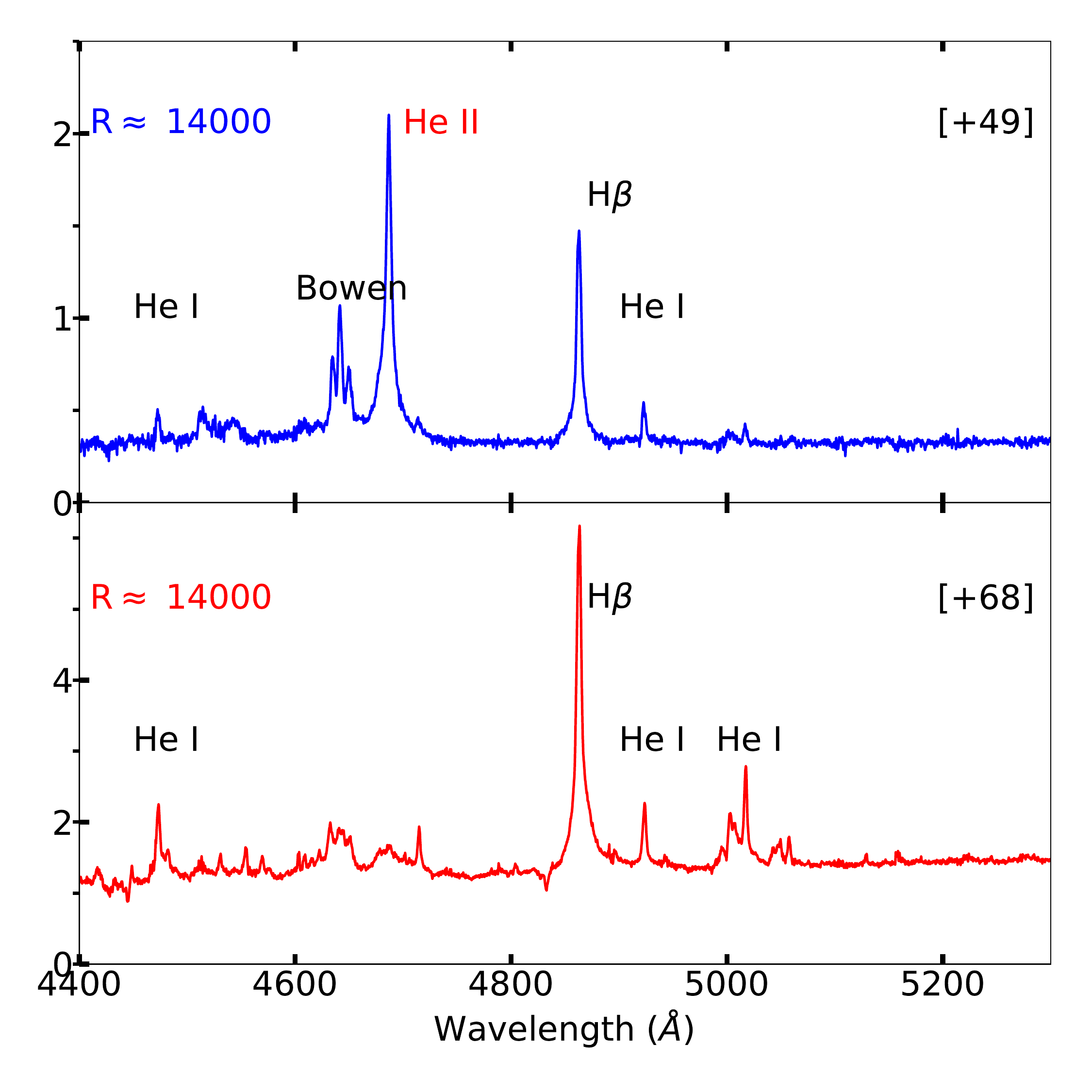}
\caption{The spectral evolution during the first two spectral epochs, but focusing on the evolution of \eal{He}{II} 4686\,\AA{} and H$\beta$ emission lines. The spectra shows the fading/disappearance of the \eal{He}{II} line and Bowen line between the early- and mid-rise.}
\label{Fig:main_spec_Hbeta}
\end{center}
\end{figure}

On day 68, the spectra of Gaia22alz show dramatic changes, with the \eal{He}{II} 4686\,\AA{} emission line almost completely disappearing (see Figure~\ref{Fig:main_spec_Hbeta} for a zoom-in on the region around \eal{He}{II} 4686\,\AA{} and H$\beta$), and the emergence of P Cygni absorption components in the Balmer and \eal{He}{I} lines at blueshifted velocities of around $-1700$\,km\,s$^{-1}$ (see the line profiles in Figure~\ref{Fig:Hbeta}). Between days 68 and 96 (the mid-rise phase), the Balmer and \eal{He}{I} P Cygni absorptions become more prominent, while showing a decrease in the radial velocities of the P Cygni absorptions to $-1200$\,km\,s$^{-1}$. At this stage, the spectra are more consistent with a classical nova eruption before peak brightness. \citet{Aydi_etal_2020b} discussed the spectral evolution of novae before and after they reach optical peak, and suggested that the deceleration observed in the P Cygni profiles during the rise to peak is an optical depth effect. 

During the late-rise phase, between days 135 to 178, the spectra show more dramatic changes with the \eal{He}{I} emission lines fading, and the \eal{Fe}{II} (42), (48), and (49) multiplets taking over. At this stage the spectra are characteristic of classical novae before peak, originating in optically thick ejecta during the so-called iron curtain phase \citep{Shore_2014}. This evolution from a spectrum dominated by Helium to \eal{Fe}{II} emission lines during the rise to peak has been observed in a few other novae (e.g., V5558~Sgr and T Pyx; \citealt{Tanaka_etal_2011_159,Izzo_etal_2012,Ederoclite_2014}) and is suggested to be due to a change in the optical depth 
of the ejecta 
(e.g., \citealt{Shore_2014,Hachisu_Kato_2022}).
The origin of the so-called He/N and \eal{Fe}{II} evolution/spectroscopic classes \citep{Williams_1992,Williams_2012,Shore_2014} is not the focus of this paper and it will be the topic of a future work (Aydi et al. 2023 in prep.). Instead, we will mainly focus on the spectral evolution during the early- and mid-rise phases. The overall spectral evolution of Gaia22alz during these early phases is comparable to that of the 2011 eruption of T Pyx (e.g., figure 3 in \citealt{Arai_etal_2015}) but on a much slower timescale. 

After the nova peaked (day $\sim$180 onwards), the spectral lines show multiple absorption features at different velocities (e.g., days 203, 212, and 236 in Figure~\ref{Fig:Hbeta}). During this phase the light curve evolution also hints to the presence of multiple maxima (Figure~1). The emergence of multiple spectral absorption components in the spectra of ``flaring novae'', correlating with the multiple peaks in the light curve, has been investigated previously (e.g., \citealt{Chochol_etal_2003,Pejcha_2009,Tanaka_etal_2011_159,Tanaka_etal_2011,Walter_2016,Harvey2018,Shore_etal_2018,Aydi_etal_2019_I,Aydi_etal_2020a,Aydi_etal_2020b}), and has been suggested to be due to either multiple phases of mass-loss (multiple ejections) or due to changes in the radius of the optical photosphere, caused by variability in the emitting central source. After day 300, the light curve shows a dramatic decline, possibly due to dust condensation. During the decline, the spectral line profiles show a broad emission (FWZI $\gtrsim$ 3000\,km\,s$^{-1}$), with slower absorption/emission superimposed on top of the broad base emission (see Figure~\ref{Fig:Hbeta} last panel; day 331). This is consistent with the spectral evolution of classical novae past peak brightness as described by \citet{Aydi_etal_2020b}.

\subsection{Reddening and distance}
\label{sec_dist}

Based on Gaia DR3, which observed Gaia22alz in quiescence (before its 2022 outburst), the system has a bias-corrected \citep{2022arXiv220800211G} parallax of $\varpi = 0.084\pm0.122$ mas. Based on this reported parallax and its uncertainty, it is not feasible to derive an accurate distance estimate to Gaia22alz, except to say that formally the distance is $\gtrsim 3.0$ kpc at the 95\% level, so very nearby distances are disfavored. This is a weak constraint but consistent with the distance derived from the larger reddening values (see below).

\begin{figure}
\begin{center}
\includegraphics[width=\columnwidth]{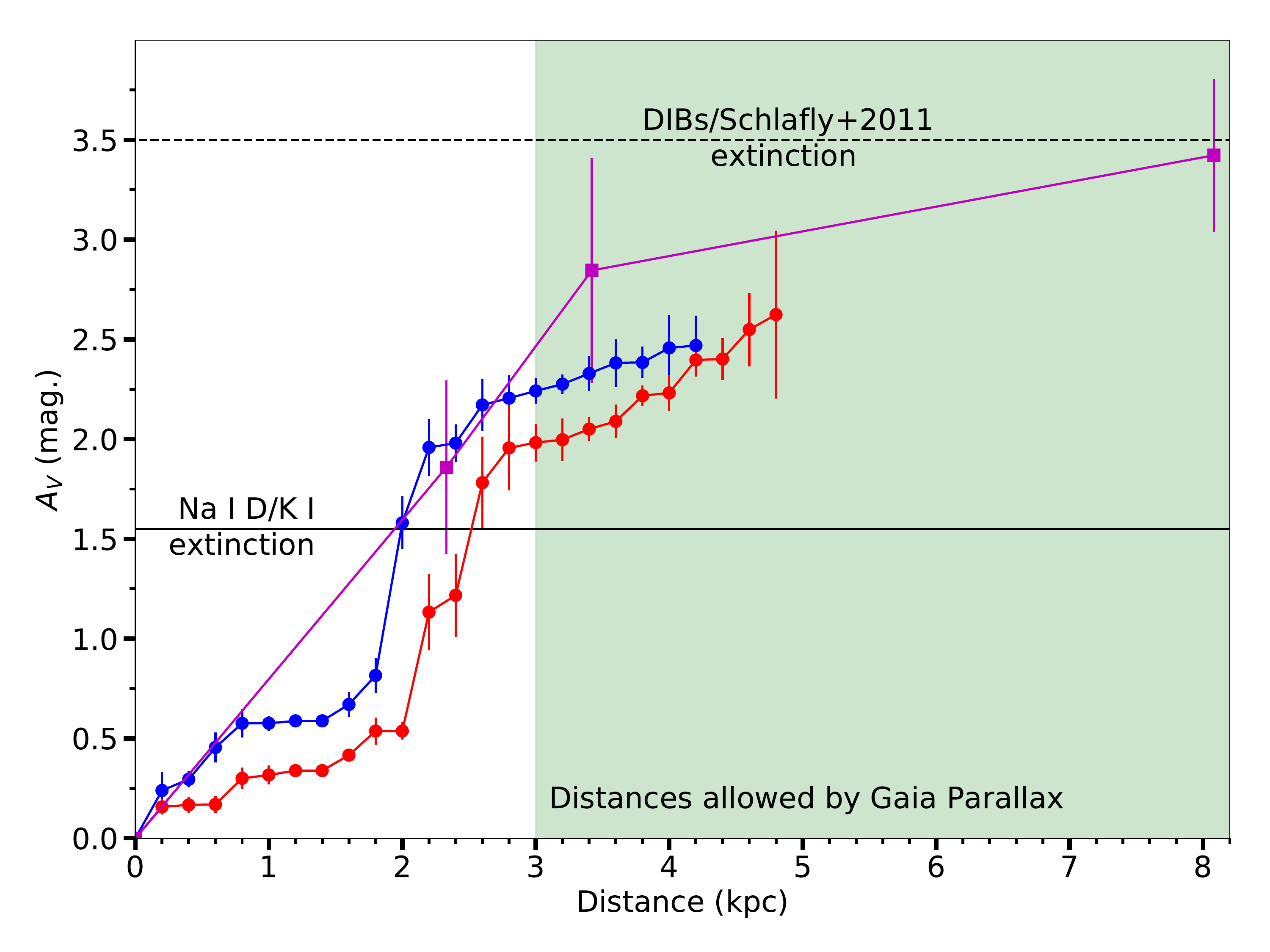}
\caption{The extinction along the line of sight to Gaia22alz plotted against the distance. We choose the closest directions to the position of Gaia22alz from the \citet{Chen_etal_2019} and \citet{Marshall_etal_2006} Galactic reddening maps. The blue and red curves are the extinction values from the \citet{Chen_etal_2019} maps in the direction ($[l,b] =[277.55^{\circ},-3.75^{\circ}]$) and ($[l,b] =[277.55^{\circ},-3.65^{\circ}]$), respectively. The magenta curve represents the extinction values from the \citet{Marshall_etal_2006} maps in the direction ($[l,b] =[277.50^{\circ},-3.75^{\circ}]$). Distances allowed by the \emph{Gaia} parallax at 95\% confidence are shaded in green ($>$3\,kpc).}
\label{Fig:reddvsdist}
\end{center}
\end{figure}

The Galactic reddening maps of \citet{Schlafly_etal_2011} estimate a total reddening value of $A_V$ = 3.47\,mag in the direction of Gaia22alz. \citet{Chen_etal_2019} use measurements from the Gaia DR2, 2MASS and WISE surveys to estimate reddening as a function of distance in the Galactic plane. Along the line of sight to Gaia22alz, they find that at around 2\;kpc there is a significant jump in the reddening towards Gaia22alz (see Figure \ref{Fig:reddvsdist}, which plots the reddening along the line of sight to Gaia22alz as a function of distance). However, the Chen et al. maps only extend to $\approx$ 4.5\,kpc in this direction. \citet{Marshall_etal_2006} use the 2MASS survey and a slightly lower resolution grid to estimate a three dimensional reddening map out to larger distances. This map extends to around 8\,kpc in the direction of Gaia22alz.

For a reddening estimate along the line of sight to Gaia22alz, which can in turn be used to constrain the distance, we use absorption lines from diffuse interstellar bands (DIBs), which are tracers of interstellar dust. We measure the Equivalent Width (EW) of the lines at 5780.5, 5797.1, 6196.0, 6204.5, and 6613.6\,\AA{} and use the relations of \citet{Friedman_etal_2011} to derive an average $E(B-V)$ = 1.13\,mag. This implies $A_V$ = 3.5\,mag, assuming a reddening law $R_V$ = 3.1. This value is consistent with the integrated reddening along this line of sight, as estimated by the \citet{Schlafly_etal_2011} Galactic reddening map. 

We also use the EW of the \eal{Na}{I} D and \eal{K}{I} interstellar lines to derive an additional reddening estimate. The EW of the \eal{Na}{I} D1 and D2 lines are 0.58 and 0.52\,\AA{}, while that of the \eal{K}{I} line at 7699 \AA\ is 0.105\,\AA{}. These interstellar lines are shallow and narrow, arguing that they are not saturated. Using these values along with the empirical relations from \citet{Munari_Zwitter_1997}, we derive $E(B-V) \approx 0.5$\,mag, translating to $A_V$ = 1.55\,mag assuming $R_V$ = 3.1. This value is half that of the \citet{Schlafly_etal_2011} Galactic reddening maps and the value we derive from the DIBs.

We use our reddening estimates and the three-dimensional Galactic reddening maps of \citet{Chen_etal_2019} to estimate the distance to Gaia22alz. In Figure~\ref{Fig:reddvsdist}, we compare reddening as a function of distance (along several lines of sight that pass close to Gaia22alz) with our reddening values derived from the DIBs and the \eal{Na}{I}D/\eal{K}{I} lines.
Since the maps of \citet{Chen_etal_2019} use measurements from the Gaia DR2, 2MASS and WISE surveys, we use the reddening laws of \citet{Wang_etal_2019} to convert $E(B-V) = 1.13$\,mag into $E(G_B-G_R) = 1.44$, $E(G-K_s) = 2.49$, and $E(H-K_s) = 0.18$\,mag. All these values yield a distance larger than 4.5\,kpc. However, if we use $E(B-V) = 0.5$\,mag (derived from the \eal{Na}{I} D and \eal{K}{I} interstellar lines), we get a distance of 2.5$\pm$0.5\,kpc using the same maps.
This lower reddening value, and correspondingly lower distance, is not consistent with the Gaia parallax, which suggests a distance likelihood in excess of 3\,kpc. 

Therefore, throughout the paper we assume a distance $d > 4.5$\,kpc and a reddening value $A_V$ = 3.5\,mag towards Gaia22alz, as implied by the DIBs and consistent with the Gaia parallax and the \citet{Schlafly_etal_2011} maps. However, we also discuss the implications of shorter distances and lower reddening towards Gaia22alz in the following sections.

\subsection{The progenitor system}
\label{sec_prog}
Archival data from the Optical Gravitational Lensing Experiment (OGLE; \citealt{Udalski_etal_2015}), taken between 2017 and 2020, measured average magnitudes $I= 17.57 \pm 0.12$ and $V=18.66 \pm 0.09$ of the progenitor system of Gaia22alz during quiescence. In archival data from the Dark Energy Camera Plane Survey (DE-CaPS; \citealt{Schlafly_etal_2018}) the source has median pre-eruption magnitudes of
$g \simeq 19.0$, $r \simeq 18.2$, $i \simeq 18.1$, $z \simeq 18.0$, and $y \simeq 17.9$. Moreover, the AB magnitudes of the progenitor system, $u=19.56\pm0.02$,  $g=19.04\pm0.01$, $r=18.40\pm0.02$, $m(H\alpha)=18.23\pm0.02$ and $
i=18.01\pm0.02$, measured by the VST Photometric H$\alpha$ Survey of the Southern Galactic Plane and Bulge (VPHAS+; \citealt{Drew_etal_2014}) on 2012-12-20 (HJD=2456282.76), indicate a hot source and no substantial H$\alpha$ emission. 

\begin{figure}
\begin{center}
\includegraphics[width=0.9\columnwidth]{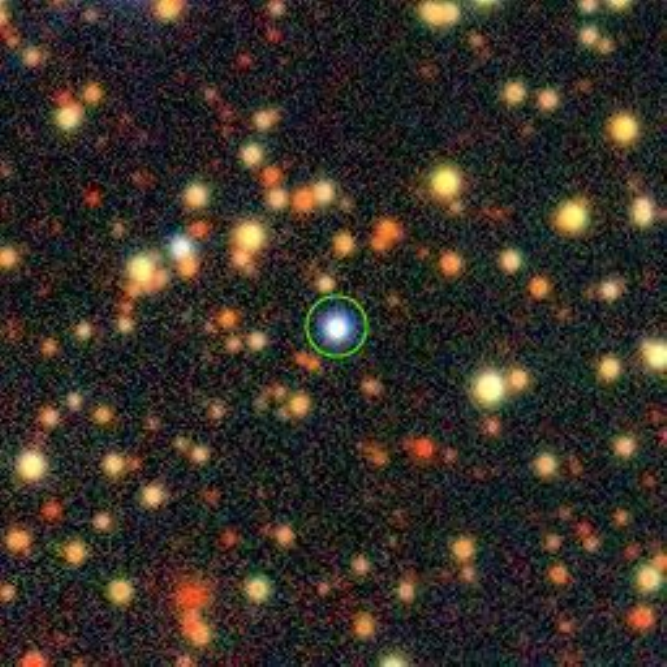}
\caption{A 1$\times$1 arcmin DE-CaPS $grz$ image of the field of Gaia22alz taken on 2017-10-04. The candidate progenitor system of Gaia22alz is circled in green.}
\label{Fig:chart}
\end{center}
\end{figure}

The above measurements imply an absolute quiescent $V$-magnitude $M_V \leq$ 1.8, assuming $A_V$ = 3.5\,mag and a distance $\geq$ 4.5\,kpc. This means that the system is luminous compared to other cataclysmic variables (CVs) in quiescence (e.g., \citealt{Warner_etal_1987MNRAS}). At such a brightness, the secondary star could be more evolved than a dwarf star, i.e., a sub-giant or even a giant. However, the colors of the system, $(g-i)_0 \simeq -0.9$ and $(V-I)_0 \simeq -1.73$ (assuming $A_V$ = 3.5\,mag), are blue for a typical CV (whether it has a dwarf or giant companion; \citealt{Szkod_1994,Bruch_etal_1994,Zwitter_Munari_1995,Szkody_etal_2011,Kato_etal_2012}), ruling out the possibility of a giant evolved secondary. In Figure~\ref{Fig:chart} we show a combined $grz$ color image of the region of Gaia22alz from the DE-CaPS survey \citep{Schlafly_etal_2018}, showing the progenitor system.

\begin{figure}
\begin{center}
\includegraphics[width=\columnwidth]{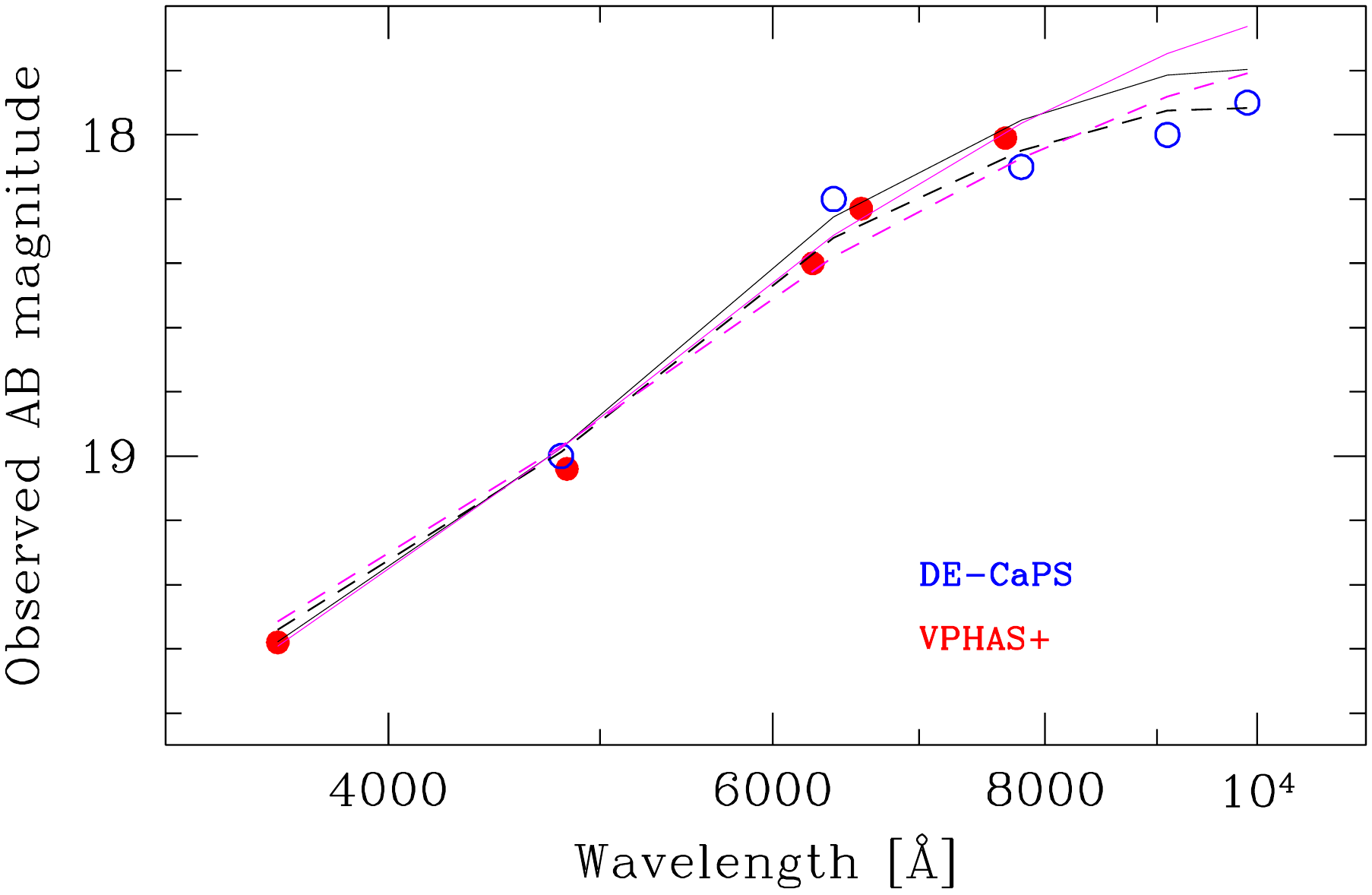}
\caption{The spectral energy distribution of the progenitor system of Gaia22alz constructed using DE-CaPS and VPHAS+ data. We fit two models to the SED: a blackbody model with Rayleigh-Jeans tail (black line) and an optically thick steady accretion disk model (magenta line) to the VPHAS+ (solid lines) and De-CaPS (dashed lines) data.}
\label{Fig:SED}
\end{center}
\end{figure}

We created a spectral energy distribution (SED) using the multi-band photometry taken during quiescence by VPHAS+ and DE-CaPS, and we successfully fit the resulting SED with a reddened Rayleigh-Jeans tail of a hot blackbody (Figure~\ref{Fig:SED}). The fit to both datasets provide almost identical outcomes.
The best fit has been obtained with a radius $R=(1.595 \pm 0.059) \times
10^{10}\, (T/10^5\,K)^{-1}\, (d/5\,{\rm kpc})$\,cm
and $E(B-V)=1.11\pm0.02$\,mag. Note that this reddening value is practically
identical to that we derived from the DIBs and the total Galactic reddening resulting from \citet{Schlafly_etal_2011} maps in the direction of Gaia22alz. There is no evidence for a cool companion, as up to $\sim$1 micron ($Y$ band) the SED is well-reproduced by the hot blackbody, and in particular, any companion
could not contribute more than a few per cent to the SED, which rules out
a yellow/red giant donor.

Integrating under the blackbody, the source is luminous with a luminosity $L=(4740\pm350)\,(T/10^5\,K)^2\,(d/5$\,kpc)$^2$ L$_{\odot}$, indicating the presence of a bright emitting source, which could be due to ongoing nuclear burning on the white dwarf surface \citep{Schaefer_Collazi_2010,Wolf_etal_2013,Zemko_etal_2015,Zemko_etal_2016,Aydi_etal_2018_2}. 
However, since a thermonuclear runaway took place on the white dwarf in 2022, leading to the nova eruption of Gaia22alz, it is unlikely that the system has been undergoing nuclear shell burning. CVs are known to show strong emission lines during quiescence \citep{Jiang_etal_2013,Breedt_etal_2014,Hou_etal_2020,Hou_etal_2023}. These emission lines could potentially affect the colors of the system during quiescence, leading to bluer colors. While based on the VPHAS+ photometry, we do not observe significant H$\alpha$ emission above the SED BB, emission lines might still have an effect on the overall emission and colors. Therefore, using a blackbody fit to describe the emission in CVs at quiescence is not necessarily an ideal assumption. Noting this caveat, we also attempt to fit the SED with an optically thick steady accretion disk model ($F_{\nu}\propto\nu^{1/3}$; Figure~\ref{Fig:SED}; magneta line). The best fit to the observed SED results with $E(B-V)=0.639 \pm 0.012$\,mag. This value is more consistent with the lower reddening values derived from the \eal{Na}{I} D and \eal{K}{I} interstellar lines. 

\begin{figure*}[!t]
\begin{center}
\includegraphics[width=\textwidth]{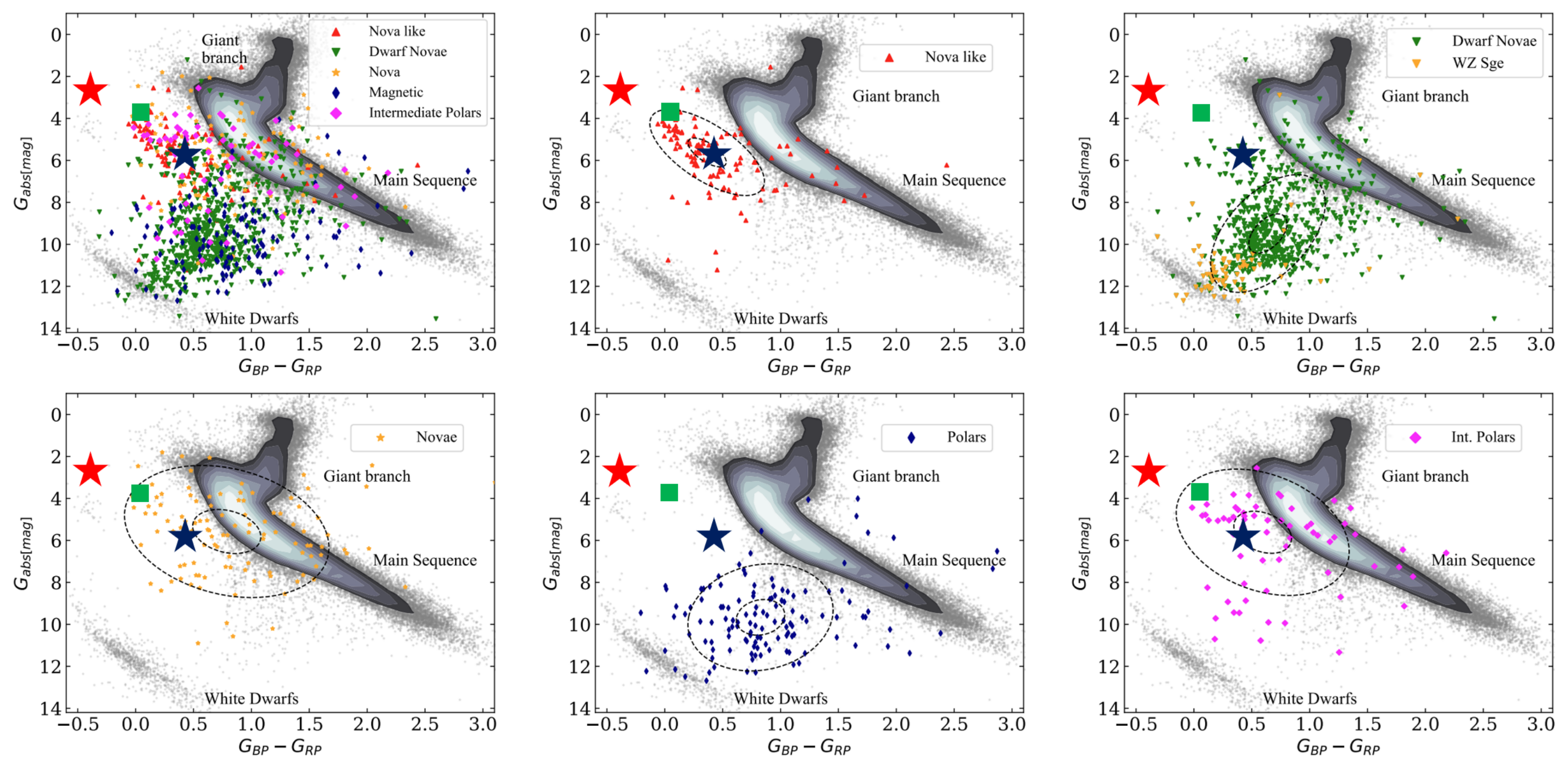}
\caption{Color-magnitude diagram adopted from \citet{Abril_etal_2020} showing the colors and absolute magnitudes, measured by Gaia, for several CV systyms from different sub-types (classical nova and dwarf nova systems in quiescence, nova-likes, polars, intermediate polars, all taken from the Catalog and Atlas of Cataclysmic Variables \citealt{Downes_etal_2001}). The dashed ellipses represent 1 and 3$\sigma$ of each sub-type bivariate Gaussian distribution (see \citealt{Abril_etal_2020} for more details). We over-plot the progenitor system of Gaia22alz at different distances; the red star corresponds to $d = 4.5$\,kpc and $A_V$ = 3.5\,mag (derived based on the DIBs), while the blue star is for $d = 2.5$\,kpc and $A_V$ = 1.55\,mag (derived based on the \eal{Na}{I} D and \eal{K}{I} lines). We also add a case of an intermediate distance/reddening (green square; $d = 3.5$\,kpc and $A_V$ = 2.5\,mag).}
\label{Fig:CMD}
\end{center}
\end{figure*}

In order to compare the brightness and colors of Gaia22alz with other CVs and to better constrain the nature of the secondary star, we place the progenitor system of Gaia22alz on a color-magnitude-diagram (CMD) adopted from \citet{Abril_etal_2020} in comparison with different CV sub-types (e.g., nova systems, dwarf novae, nova-likes, and magnetic CVs; Figure~\ref{Fig:CMD}). The colors and magnitudes of these CV systems are all Gaia measurements. At a distance of 4.5\,kpc and $A_V$ = 3.5\,mag (red star in Figure~\ref{Fig:CMD}), the system is much bluer than other CV systems and is relatively brighter. However, for a distance of 2.5\,kpc and $A_V$ = 1.55\,mag (blue star in Figure~\ref{Fig:CMD}), the system has comparable colors and brightness to other CV systems, particularly novae and nova-like systems. Both cases, small or large distance/reddening, still rule out the presence of a giant, evolved secondary. It is possible that the distance and reddening derived from the DIBs are overestimated. Therefore, intermediate distances and reddening values (e.g., $d$ = 3.5\,kpc and $A_V = 2.5$; green square in Figure~\ref{Fig:CMD}), could bridge the gap between the colors/brightness of Gaia22alz and other CVs.

In summary, the Gaia parallax of the progenitor system of Gaia22alz favors larger distances ($d>$ 3\,kpc) and therefore larger reddening towards the system. However, at such large distances and high reddening values, the progenitor system is relatively luminous and characterized by significantly bluer colors compared to other CVs, as shown in Figure~\ref{Fig:CMD}. This might lead us to suggest that the lower reddening/shorter distances are more reasonable to explain the colors and brightness of the system during quiescence, despite them being contradictory with the Gaia parallax. However, at such a distance and reddening, the absolute peak magnitude of Gaia22alz during eruption would be just $M_V = -2.7$ (assuming $d$=2.5\,kpc and $A_V = 1.55$), which is much smaller than typical peak brightness of novae ($M_V \approx -5$ to $-10$; \citealt{Shafter_2017,Shara_etal_2017_apr}). Could this mean that Gaia22alz is not a classical nova? We further elaborate on these speculations in the following sections. 
It is also possible that the actual distance and reddening towards Gaia22alz is intermediate between the values we derive from the DIBs and the ones we derive from the \eal{Na}{I} D and \eal{K}{I} lines.

\section{Discussion}
\label{Disc}
\subsection{Detecting the visible signature of the UV/X-ray flash?}

Novae observed during the rise to peak brightness typically show spectra dominated by P Cygni lines of \eal{H}{I} and either \eal{Fe}{II} or He/N, originating in the dense nova ejecta (e.g., \citealt{McLaughlin_1944,McLaughlin_1947,Payne-Gaposchkin_1957}). The line profiles are characterized by velocities ranging from a few hundreds km\,s$^{-1}$ \citep{Aydi_etal_2020b} up to a few thousands km\,s$^{-1}$ in some extreme cases (e.g., the very fast nova V1674~Her; \citealt{ATel_14710}). \citet{Aydi_etal_2020b} observed a large sample of novae before they reached optical peak, and found that all of them show the same evolution. Particularly, they found that the optical spectra before peak are dominated by P Cygni profiles originating in the dense nova outflow. Gaia22alz is an exception. The first two spectra obtained on days 45 and 49 were not typical of classical novae observed before peak. More precisely, the absence of P Cygni absorption, the relatively narrow emission lines (FWHM $\approx$ 400\,km\,s$^{-1}$), and the strong \eal{He}{II} and \eal{C}{IV} emission lines are unusual for a classical nova pre-maximum spectrum. These features, observed at a stage when the eruption was less than four magnitudes above quiescence, led \citet{ATel_15270} to suggest that Gaia22alz could be an unusual disk instability dwarf nova event or something more exotic. Here we discuss the origin of these features and speculate about the reasons for the gradual rise in the optical light curve of nova Gaia22alz. 

After the onset of the thermonuclear runaway, the CNO chain reactions become the main source of energy generation in a nova, producing radioactive elements which decay ($\beta^{+}$-decay), releasing a large amount of energy. It has been theorized \citep{Starrfield_etal_1990,Krautter_2002,Hillman_etal_2014} and now observationally confirmed \citep{Konig_etal_2022} that the dramatic increase in fusion rate is followed by a short X-ray (and even UV) flash, before the dramatic increase in the visible brightness. The X-ray/UV flash terminates when the nova envelope has had time to expand, thereby cooling down rapidly and becoming optically thick to the emission from the hot white dwarf.
The peak of the nova's SED shifts into the visible band, turning the nova into an optical transient. 
The optical spectrum of the nova during this phase (rise to visible peak) is typically dominated by P Cygni profiles \citep{Aydi_etal_2020b}. However, it is not clear how the visible emission from the nova manifests during the X-ray/UV flash phase.

The first two spectra we obtained for Gaia22alz possibly capture the late stages of an extended, slow X-ray/UV flash. Unfortunately, we do not have any X-ray or UV observations taken during the rise to peak. However, the strong high-excitation \eal{He}{II} 4686\,\AA{} and \eal{C}{IV} emission lines in the first two spectra imply emission from 
a hot ionized gas (a few times $10^{5}$\,K; e.g., \citealt{Morisset_Pequignot_1996,Bonning_etal_2013}). Below, we suggest two possible origins for the early, gradual optical rise and the relatively narrow, high ionization spectral features we observe in the early optical spectra:

(1) The optical emission is originating in the outer layers of the (yet optically thin) nova envelope, in a scenario where the bulk of the envelope has not yet reached the escape velocity and has not expanded enough in size to cause a bright optical transient (i.e., a transient characterized by an amplitude of more than 8 magnitudes). The outer layers of this slightly puffed-up nova envelope reprocess the high-energy X-ray/UV emission from the white dwarf surface, resulting in the spectral features we observe in the first 50 days and the early brightness increase. This means that the four magnitudes of gradually increasing optical brightness during the first 50 days might be taking place during the X-ray/UV flash. While the X-ray/UV flash is often predicted to be short (of the order of a few hours; \citealt{Starrfield_etal_1990,Krautter_2002,Konig_etal_2022}), \citet{Hillman_etal_2014} modeled the precursor UV flash for a range of system parameters, and found that for some models (e.g., a low mass white dwarf with $M_{WD}$ = 0.65\,M$_{\odot}$ and
a core temperature $T_c=10^7$\,K, accreting at a rate $\dot{M}$= $10^{-10}$\,M$_{\odot}$\,yr$^{-1}$), a UV flash preceding the optical peak emission can last for up to 60 days. This timescale is consistent with the duration of the early rise in Gaia22alz ($\approx$ 50 days). Moreover, based on the slow evolution of the light curve, we indeed expect that the WD mass in Gaia22alz is lower than 1 solar mass.

The wings of the Balmer lines, which extend to velocities $\gtrsim$2000\,km\,s$^{-1}$ at this early epoch (see Figure~\ref{Fig:Hbeta}), possibly originate in a low-density, radiation driven wind from the white dwarf surface, before the bulk of the envelope is ejected. 

After the X-ray/UV-flash and the early visible rise, we expect that the bulk of the nova envelope is ejected, possibly aided by the binary motion \citep{Chomiuk_etal_2014}. This would lead to the final climb towards peak brightness (days $\sim$ 50 to 180). During this stage the envelope cools to the critical temperature at which a recombination wave forms, so the envelope becomes opaque to the UV emission \citep{Shore_2014}. This manifests as a transition in the optical spectra from a spectrum dominated by narrow emission lines to one dominated by higher velocity P Cygni profiles. This also manifests as the sudden disappearance of the high excitation lines of \eal{He}{II} and \eal{C}{IV} (see first two panels in Figure~\ref{Fig:main_spec}). This is similar to the overall spectral evolution observed during the 2011 eruption of T Pyx (\citealt{Arai_etal_2015}; see also Table~\ref{table:comparison}). T Pyx also shares other similarities with Gaia22alz, particularly the optical light curve evolution which showed a relatively slow rise (40 days) to visible brightness with a plateau phase at peak that lasted for more than a month \citep{Schaefer_etal_2013}. 

Optical multi-band photometry taken during the first 100 days would have been very informative to check if there is any change in the temperature of the emitting source and to derive its size. However, reported AAVSO observations only start after day 97, missing the crucial early- and mid-rise phases. Nevertheless, if the temperature of the emitting source is of the order of a few times 10$^{5}$\,K (as implied by the high-excitation emission lines) and the brightness is $g \approx$ 15 (implying $M_g \approx$2.5\,mag for a distance of 4.5\,kpc), the optical emission should be originating close to the surface of the white dwarf (at $r\sim10^{9}$\,cm; an emitting region  comparable to the radius of the white dwarf).

(2) Another possibility is that the optical emission is originating in the accretion disk, which is reprocessing high-energy (X-ray and/or UV) emission from the white dwarf surface. Similar to the scenario described above, the first 50 days represent the optical signature of an extended X-ray/UV flash, but in this case the visible emission is originating in the disk rather than the outer layers of a slightly puffed-up nova envelope. Emission lines of Balmer, \eal{He}{I}, and \eal{He}{II} are often observed in CV systems and dwarf nova outbursts (e.g., \citealt{Morales-Rueda_Marsh_2002}), and they are characterized by velocities of a few hundreds km\,s$^{-1}$ --- similar to the ones we observe in the first two spectra of Gaia22alz. Could reprocessed emission by the accretion disk lead to a visible brightness increase of $\sim$4 magnitudes and absolute optical magnitudes spanning +2 to $-2.5$ (assuming $d$=4.5\,kpc and $A_V$ = 3.5\,mag), during the first 50 days of the outburst? In the case of supersoft X-ray sources (e.g., \citealt{vandenHeuvel_etal_1992}), which are similarly powered by nuclear burning on the white dwarf and which part of their visible emission is associated with emission from the accretion disk (in addition to the companion star), we observe absolute magnitudes between +1.5 and $-2.5$ (e.g., \citealt{Smale_etal_1988}), comparable to the brightness of Gaia22alz during the first 50 days of the eruption. However, this is merely an indirect argument that visible emission from the accretion disk could potentially explain the early rise of Gaia22alz.

Whether the high-energy emission is reprocessed by the outer layers of the still-optically-thin and hot nova envelope or the accretion disk, Gaia22alz could be one of the rare times we observe the optical signature of the hot X-ray/UV flash, which is often missed in other novae. We can speculate that the extremely slow and gradual rise of nova Gaia22alz and the diverse all-sky surveys that are observing the sky more efficiently at deep magnitudes (visible magnitudes down to 17--18\,mag) is what allowed us to catch this early phase of a nova eruption. 
Given the lack of multi-wavelength observations during this early and critical phase, which prevent us from drawing definitive conclusions, we consider alternative scenarios below and explore their pros and cons for explaining the unusual observational features of Gaia22alz.

\begin{table*}
\centering
\caption{A comparison between the parameters of Gaia22alz (for different distance and extinction values) with the properties of classical novae and dwarf novae, along with the 2005 and 2019 outbursts of V1047~Cen \citep{Aydi_etal_2022}, and the 2011 eruption of T~Pyx \citep{Arai_etal_2015,Godon_etal_2018}.}
\begin{tabular}{lccccc}
\hline
System & $\Delta m$ $^{a}$& $M_{V,\mathrm{peak}}$ & \eal{Fe}{II} P Cygni $^{b}$& \eal{He}{II}/\eal{C}{VI} & Dust formation? \\
 & [mag] & & profiles & during the rise & \\
\hline
Gaia22alz ($d$= 4.5\,kpc; $A_V$ = 3.5)& 7.8 & $-5.9$ & Yes & Yes & Yes\\
Gaia22alz ($d$= 2.5\,kpc; $A_V$ = 1.55)& 7.8  & $-2.7$ & Yes  & Yes & Yes \\
Classical novae & 8--15 & $-5$ to $-11$ & Yes & No & Yes\\
Dwarf novae & 2--5 & $>$ 0.0 & No & \eal{He}{II} Yes & No\\
V1047~Cen (2019) & 6 & $-2.0$ & No & Yes & No\\
V1047~Cen (2005; classical nova) & 13 & $-8.0$ & Yes & No & No data\\
T Pyxidis (2011; classical nova) & 9 &  $-8.1$ & Yes & Yes & No\\
\hline
\end{tabular}
\label{table:comparison}
\raggedright {$a${$\Delta m$ is is outburst amplitude.}\\
$b${Appearance of prominent \eal{Fe}{II} P Cygni profiles near optical peak.}}
\end{table*}

\subsection{Alternative explanations} \label{sec:alt}
Below we suggest some alternative explanations to the early spectral features of Gaia22alz and its unusual, gradual rise. The characteristics of some of these different scenarios are compared in Table \ref{table:comparison}, considering two different distance and reddening assumptions for Gaia22alz.
\begin{itemize}

    \item \textbf{Shock interaction from an early mass-loss phase?}
    The main alternative scenario we suggest here is that, during the first weeks of the eruption (before the bulk of the nova envelope is ejected), the system loses mass from the L2 outer Lagrangian point. The gravitational potential of the rotating binary is not axially symmetric, so the matter leaving from L2 exhibits small fluctuations in its radial velocity (see \citealt{Pejcha_etal_2016} for a detailed description of the model). This causes the matter being ejected from L2 as spiral arms to collide with itself at around 10 times the semi-major axis of the binary, creating stationary shocks that radiate some of the kinetic energy of the L2 outflow. In Section~\ref{appC} of the Appendix, we present calculations for the temperature of the shocked gas and the luminosity of the line emission from these stationary shocks. Our calculations show that the shocks would heat the gas to a temperature of $\sim$few times 10$^5$\,K with emission line luminosity of 10$^{35}$\,erg\,s$^{-1}$, consistent with the early spectra of Gaia22alz.

    \item \textbf{A dwarf nova outburst preceding/triggering a classical nova?} The features detected in the first two spectral epochs of Gaia22alz resemble features observed in disk instability dwarf nova outbursts (see \citealt{Warner_1995} for a review). During outburst, the spectrum of a dwarf nova is typically dominated by absorption features with fill-in by emission lines.  Moreover, the spectra could exhibit strong emission lines, particularly at later stages of the outburst (e.g., \citealt{Bruch_1989,Bruch_Schimpke_1992,Morales-Rueda_Marsh_2002,Han_etal_2020}). Strong emission lines of \eal{He}{II} at 4686\,\AA{} are also commonly observed in the spectra of dwarf novae \citep{Han_etal_2020}. The gradual increase of just a few magnitudes in the optical light curve during the first 50 days of Gaia22alz eruption is consistent with the amplitude of a dwarf nova outbursts ($\approx$ 2 to 5 mag). Therefore, could it be that the classical nova eruption of Gaia22alz was coincidentally preceded or maybe even triggered by a dwarf nova outburst? 
    \citet{mroz16} observed multiple dwarf nova outbursts preceding the classical nova eruption of nova V1213~Cen in 2009. Similarly, V392~Per and V1017~Sgr showed DN outbursts two and eight years, respectively, prior to their classical nova eruptions \citep{Murphy-Glaysher_etal_2022,Salazar_etal_2017}. Meaning that DN outburst could precede a classical nova eruption by a few years, and potentially in the case of Gaia22alz, happening around the same time as the classical nova eruption started. Could this be the case? While this is possibility, it is unlikely that a disk instability event would be the trigger for the classical nova eruption. In particular, we note that the onset of the thermonuclear runaway is expected to take place years before the classical nova becomes a visible transient \citep{Starrfield_etal_1974,Prialnik_1986,Shen_Quataert_2022}. Moreover, the material dumped onto the white dwarf from a dwarf nova outburst is suggested to be of the order of 10$^{24}$\,g ($\sim 10^{-9}$\,M$_{\odot}$) in systems with dwarf companions (e.g., \citealt{Cannizzo_etal_1993}), meaning that the total accreted mass due to the dwarf nova outburst is likely negligible compared to the total mass of the classical nova accreted/ejected envelope (10$^{-7}$ -- 10$^{-3}$\,M$_{\odot}$; \citealt{Yaron_etal_2005}). In conclusion, it is unlikely that the eruption of Gaia22alz is triggered by a dwarf nova outburst. However there remains a slight chance of coincidence in Gaia22alz: in this scenario, the early rise of the light curve and the early spectral features (narrow emission lines of Balmer, \eal{He}{I}, and \eal{He}{II}) are due to a disk instability event which coincidentally directly precedes the classical nova eruption. Nevertheless, this possibility also does not agree well with the pre-eruption magnitudes and SED fitting (Figure~\ref{Fig:SED}). The SED fitting to the VPHAS+ and DE-CaPS data implies a steady optically thick accretion disk for at least the epochs covered by these observations. However, a steady optically thick accretion disk is inconsistent with a dwarf nova because their quiescent accretion disk are usually optically thin \citep{Han_etal_2020}. Similarly the colors and brightness of the system during quiescence are different from most DN systems (Figure~\ref{Fig:CMD}).
    
    \item \textbf{An outburst similar to the 2019 outburst of V1047~Cen?}
    Gaia22alz share similarities with the peculiar 2019 outburst of V1047~Cen, which was first identified as an outbursting classical nova in 2005 
    \citep{Aydi_etal_2022}. The 2019 outburst was characterized by a long rise (lasting more than 2 months), a slowly evolving light curve (with an outburst duration $>$400 days), and showed significant differences with the 2005 classical nova eruption experienced by the same system (Figure~1 in \citealt{Aydi_etal_2022}). In Figure~\ref{Fig:LC_comp} we show a comparison of the light curves of Gaia22alz and V1047~Cen, where both show a slow rise and continuous variability near peak brightness for several months. The spectra of the 2019 outburst of V1047~Cen also showed narrow emission lines (FWHM of a few hundred km\,$s^{-1}$) of Balmer and \eal{He}{I}, \eal{He}{II}, and \eal{C}{IV} during the rise to peak. As in Gaia22alz, \citet{Aydi_etal_2022} noted that these high-excitation \eal{He}{II} and \eal{C}{IV} emission lines disappeared as the outburst evolved. The colors of the system prior to the 2019 outburst were also significantly blue ($[V-I]_0 = -1.19$), which led \citet{Aydi_etal_2022} 
    to suggest that ongoing nuclear burning was taking place on the surface of the white dwarf. \citet{Aydi_etal_2022} suggested that the outburst of V1047~Cen may have started with a brightening of the disk due to enhanced mass transfer or disk instability, possibly leading to enhanced nuclear shell burning on the white dwarf, which was already experiencing some level of quasi-steady shell burning in the wake of the 2005 nova. The nuclear shell burning eventually led to the generation of a wind/outflow, resulting in significant broadening in the emission lines (see \citealt{Aydi_etal_2022} for more details). 
    
    Could Gaia22alz's outburst be similar to that of the 2019 outburst of V1047~Cen, rather than being a classical nova? While there are some similarities between V1047~Cen and Gaia22alz (e.g., the slow rise to peak brightness, the high-excitation lines during the rise to visible peak, the blue colors of the progenitors, and the slowly evolving outbursts), the two events  also show discrepancies. As the event evolved, the 2019 outburst of V1047~Cen did not show typical observational features of classical novae, while Gaia22alz did develop such features (e.g., peak absolute magnitude and emergence of \eal{Fe}{II} P Cygni profiles, potential dust formation). As detailed in Table~\ref{table:comparison}, at a distance of 4.5\,kpc and for the relatively large extinction value, Gaia22alz would have a peak absolute magnitude of $M_V = -5.9$, consistent with a classical nova eruption. However, V1047~Cen only reached $M_V \approx -2$ in 2019 (much smaller than typical brightness of classical novae). The gap is bridged if we consider shorter distances and lower extinction towards Gaia22alz. Nevertheless, the optical spectral evolution of the two systems is still different, with Gaia22alz exhibiting spectra dominated by strong Balmer and \eal{Fe}{II} P Cygni profiles as the outburst evolved, which was not the case of V1047~Cen, i.e., no prominent \eal{Fe}{II} P Cygni profiles. Note that the 2005 classical nova eruption of V1047~Cen, showed typical spectral features of classical novae with spectra dominated by P Cygni profiles of Balmer and \eal{Fe}{II}  near optical peak; \citealt{Walter_etal_2012,Aydi_etal_2022}. 

    \item \textbf{Pre-existing Circumbinary material?} Could the early spectral features observed in Gaia22alz originate in circumbinary material reprocessing the high-energy emission from the white dwarf surface? In this scenario, we still require high-energy luminous emission from the X-ray/UV flash, but it is then absorbed by the neutral gas in the circumbinary medium and re-emitted in the visible band. Many nova systems, particularly ones with evolved giant secondaries, are known to have dense circumbinary material enriched by the mass-losing, giant companion (e.g., RS~Oph- \citealt{Obrien_etal_2006,Booth_etal_2016}; V3890~Sgr- \citealt{Harrison_etal_1993,Orio_etal_2020,Mikoljewska_etal_2021}; V407~Cyg- \citealt{Munari_etal_2011_V407cyg}). In contrast, novae with main-sequence, dwarf secondaries are characterized by low-density circumbinary media  (e.g., \citealt{Bode_etal_1987}). The dense circumbinary
    material in the case of novae with giant secondaries manifests as low-velocity ($\approx$ 10--50 km\,s$^{-1}$) P Cygni profiles in their optical spectra (e.g., \citealt{Dobrzycka_etal_1994,Brandi_etal_2009,Mondal_etal_2018}). These velocities reflect the relatively slow mass-loss from the red giant companion \citep{Bowen_1988}. 
    
    The optical spectra of Gaia22alz do not show such features. The  
    relatively narrow emission lines observed in its early spectra are characterized by FWHM $\approx$ 400\,km\,s$^{-1}$, which is significantly faster than the velocities typically observed for winds/mass-loss from evolved stars in symbiotic systems \citep{Brandi_etal_2009,Munari_etal_2011}.  
    Moreover, the blue color of the system during quiescence argues against an evolved secondary star (see Section~\ref{sec_prog} and Figure~\ref{Fig:CMD}). Therefore, it is unlikely that the early spectra of Gaia22alz originate in circumbinary material reprocessing the emission from the nuclear burning on the white dwarf.

    \item \textbf{An unusual symbiotic nova?}
    Symbiotic novae are thermonuclear runaway events that take place on accreting white dwarfs in wide binary systems (e.g., \citealt{Kenyon_Turan_1983,Mikolajewska_2010}). These systems consist of a white dwarf accreting material from an evolved, red giant companion \citep{Belczynski_2000,Mikolajewska_2007}.
    Traditional symbiotic novae are characterized by slow outbursts that last years and even decades (in contrast with recurrent novae in symbiotic systems, which tend to evolve rapidly; see \citealt{Mikolajewska_2010}).
    The orbital periods of such systems are much longer than those of classical nova systems ($P_{\mathrm{orb}} > 100$\,days for symbiotic nova systems, compared to $P_{\mathrm{orb}} < 1$\,day for classical nova systems), and the white dwarf hosts are characterized by low masses (0.4--0.6\,M$_{\odot}$; \citealt{Mikolajewska_2003}).  
    
    While the outburst of Gaia22alz is slow compared to classical novae ($t_2$ = 212 days, qualifying it as a ``very slow" nova; \citealt{Payne-Gaposchkin_1957}), the timescales on which symbiotic novae evolve tend to be even slower (a few years up to decades; e.g., \citealt{Chochol_etal_2003,Mikolajewska_2010}). Could Gaia22alz be an unusually rapidly evolving symbiotic nova? 
    There is generally a scarcity of spectroscopic follow-up of slow symbiotic novae during their rise to optical peak to compare with Gaia22alz early spectra, but \citet{Munari_etal_2009} observed the slow symbiotic nova V4368~Sgr across 16 years of its outburst that started in 1993. The pre-maximum spectra show relatively narrow emission lines of Balmer with no P Cygni profiles, similar to the early spectra of Gaia22alz. However, they do not detect high-ionization \eal{He}{II} or \eal{C}{IV} lines. The spectral features of V4368~Sgr remained similar after its optical peak, with no P Cygni profiles emerging. This led \citet{Munari_etal_2009} to suggest that up until the time of their study, mass-loss from the nova was still very low or even absent. While the spectral evolution of V4368~Sgr might not necessarily represent all slow symbiotic nova, its overall evolution is different from Gaia22alz, which clearly showed evidence for ejected material and mass-loss as the eruption evolved. Moreover, based on the archival data of the progenitor system (see Section~\ref{sec_prog} and Figure~\ref{Fig:CMD}), it is unlikely that the companion star is a red/yellow giant, arguing against a symbiotic nova possibility.

    \item \textbf{A stellar merger?} One alternative suggestion to explain the unusual spectral and photometric evolution of Gaia22alz is that the eruption is more exotic than a nova outburst. While the spectra observed after day 50 become consistent with a classical nova eruption, the amplitude of the eruption ($\approx$8\,mag) is on the lower end of eruption amplitudes in novae \citep{Warner_2008, Kawash_etal_2021a}. The optical light curve of Gaia22alz resembles the optical light curves of stellar mergers (Luminous Red Novae; see Figure 6 in \citealt{Pastorello_etal_2019}), showing a relatively slow evolution and secondary peaks. Could it be that Gaia22alz is a stellar merger? The current dataset does not support this hypothesis; in particular, the spectra never showed a red continuum nor particular absorption bands that are usually seen in luminous red novae (e.g., TiO and VO; \citealt{2015ApJ...805L..18W,Pastorello_etal_2019}). Nevertheless, further follow-up are encouraged to confirm the nature of this event, and obtain the parameters of the binary after the system has returned to quiescence. 
\end{itemize}

\subsection{The progenitor brightness and colors}
Photometry of Gaia22alz's progenitor system during quiescence indicates emission from a luminous and hot, blue source at the large distance and extinction estimates ($d=4.5$\,kpc and $A_V$ = 3.5; see Section~\ref{sec_prog} and Figure~\ref{Fig:CMD}). Here we speculate about potential causes for this unusually high luminosity and remarkably blue colors. 

To first order, ongoing nuclear burning on the surface of the white dwarf seems to be the most reasonable explanation for the brightness and colors of the progenitor. Ongoing nuclear burning could be due to a relatively high mass transfer rate (10$^{-8}$ to 10$^{-7}$\,M$_{\odot}$\,yr$^{-1}$; e.g., \citealt{Wolf_etal_2013}). However, sustained nuclear shell burning on the white dwarf could not directly precede a 2022 thermonuclear nova eruption \citep{Starrfield_1989_bode,Wolf_etal_2013}. We can, in turn, speculate that the 2022 eruption of Gaia22alz is not a thermonuclear nova eruption, but rather a dramatic brightness increase due to wind/shell ejection driven by the nuclear shell burning, particularly due to the extremely slow and gradual rise --- in a scenario similar to that suggested for the 2019 outburst of V1047~Cen \citep{Aydi_etal_2022}, as discussed above. However, we have showed in Section~\ref{sec:alt} that there are differences in the overall evolution of Gaia22alz and V1047~Cen.

Could the accretion disk be undergoing an active state prior to the nova eruption? Several recurrent novae are shown to go through active high-accretion states prior to a nova eruption. This manifests as an increase in the system brightness by a fraction of a magnitude accompanied with brightness flickering, a change in the colors towards the blue (in the optical), and appearance of high excitation emission lines, such as \eal{He}{II}, in the optical spectra (e.g., \citealt{Munari_etal_2016}). This active state is suggested to be due to an in increase in the accretion rate onto the white dwarf, with the flickering originating in the accretion disk \citep{Luna_etal_2018}. While an increased accretion rate leading to the 2022 eruption of Gaia22alz could increase the brightness of the system, the lack of multi-wavelength observations during quiescence prevents us from reaching a conclusive explanation for the brightness/colors of the system. 

Similarities with T Pyx in quiescence: we have discussed above the similarities in the spectral and light curve evolution of Gaia22alz and T Pyx. The two systems also share similarities during quiescence. The absolute magnitude of T Pyx during quiescence is $M_V = 1.8 \pm 0.8$, assuming a distance between 2.5 and 4.5\,kpc and $E(B-V)$ = 0.35 (see, e.g., \citealt{Patterson_etal_1998,Godon_etal_2018}). This is also signifincatly bright compared to other CVs with dwarf companions and is equal to the brightness of Gaia22alz during quiescence ($M_V$ = 1.8; for $d$ = 4.5\,kpc and $A_V$ = 3.5\,mag). The colors of T Pyx during quiescence are also bluer than typical CVs, with $(B-V)_0 = -0.35$. This means that both systems are significantly bright and blue in comparison to other nova systems in quiescence, which might imply certain physical mechanisms or system properties that are responsible for the quiescent characteristics of systems like Gaia22alz and T Pyx. However, investigating these mechanisms are beyond the scope of this paper. 

A lower reddening and distance estimate? The reddening we derive from the \eal{Na}{I} D and \eal{K}{I} interstellar lines ($E(B-V)$ = 0.5) suggests a smaller distance towards Gaia22alz. At these distances and reddening, the colors of the progenitor system are consistent with other CVs in quiescence (see Figure~\ref{Fig:CMD}). 
However, as mentioned above, at such distances and reddening, the peak absolute magnitude of the nova would be $M_V \approx$ $-2.7$. This is much fainter than the peak intrinsic brightness of typical classical novae \citep{Shafter_2017}. Could this be an indication that Gaia22alz is an unusually faint classical nova (particularly due to the unusual spectral evolution and very slow and gradual rise of the eruption)? While unlikely, this would make Gaia22alz one of the least luminous classical novae on record. 

As discussed above, Intermediate values of reddening and distance (e.g., $d$=3.5\,kpc and $A_V=$2.5\,mag) could potentially bridge the gaps and result with more reasonable quiescent colors [$(V-I) = -0.2$ \& $(G_{BP} - G_{RP}) = 0.037$; see Figure~\ref{Fig:CMD}] and acceptable peak absolute magnitude ($M_{V,\mathrm{peak}} \approx -4.4$).

\section{Conclusions}
\label{sec_conc}
We present a detailed analysis of the spectral evolution of the slow nova Gaia22alz during its rise to peak brightness. The spectra obtained during the first 50 days were inconsistent with typical classical nova spectra, showing relatively narrow (FWHM $\approx$ 400\,km\,s$^{-1}$) emission lines and strong high-excitation lines of \eal{He}{II} and \eal{C}{IV}. 
However, as the eruption progressed, the spectra underwent dramatic changes, becoming more akin to classical novae. 
Based on the features observed in the early spectra, we suggest that the early optical emission originates in the outer layers of a slightly puffed-up (but not yet ejected) optically thin nova envelope or in the accretion disk, 
which are reprocessing high-energy (UV/X-ray) emission from the white dwarf surface during the so-called X-ray/UV flash phase. 
The spectral line profiles also suggest the presence of a fast wind, possibly driven by radiation of the nuclear burning white dwarf,  
before the bulk of the nova envelope is ejected. If confirmed, this would be one of the rare incidents we have observed the optical signature of a nova during the X-ray/UV flash. Thanks to ongoing all-sky surveys, we anticipate more slow classical novae to be detected at early stages of eruption, which will help us to better understand and constrain the physics of the poorly-observed X-ray/UV flash phases in classical novae.

The lack of multi-wavelength observations prevent us from confirming these suggestions, and we speculate about alternative explanations/scenarios to explain the nature of this event and its unusual observational features. Follow-up observations in the future might allow us to constrain the parameters of the progenitor system and its physical characteristics, which could help us gain insight into the nature of the eruption of Gaia22alz.

\section*{Acknowledgments}

We thank the AAVSO observers from around the world who contributed their magnitude measurements to the AAVSO International Database used in this work.

E.A. acknowledges support by NASA through the NASA Hubble Fellowship grant HST-HF2-51501.001-A awarded by the Space Telescope Science Institute, which is operated by the Association of Universities for Research in Astronomy, Inc., for NASA, under contract NAS5-26555. EA, LC, and KVS acknowledge NSF awards AST-1751874 and AST-2107070, NASA award 11-Fermi 80NSSC18K1746, and a Cottrell fellowship of the Research Corporation. JS was supported by the Packard Foundation. DAHB gratefully acknowledges the receipt of research grants from the National Research Foundation (NRF) of South Africa.
JM was supported by the National Science Centre, Poland, grant OPUS 2017/27/B/ST9/01940.
A part of this work is based on observations made with the Southern African Large Telescope (SALT), with the Large Science Programme on transients 2021-2-LSP-001 (PI: DAHB). Polish participation in SALT is funded by grant No. MEiN 2021/WK/01. This paper was partially based on observations obtained at the Southern Astrophysical Research (SOAR) telescope, which is a joint project of the Minist\'{e}rio da Ci\^{e}ncia, Tecnologia e Inova\c{c}\~{o}es (MCTI/LNA) do Brasil, the US National Science Foundation's NOIRLab, the University of North Carolina at Chapel Hill (UNC), and Michigan State University (MSU).
Analysis made significant use of \textsc{python} 3.7.4, and the associated packages \textsc{numpy}, \textsc{matplotlib}, \textsc{seaborn}, \textsc{scipy}. 
Data reduction made significant use of \textsc{MIDAS FEROS \citep{Stahl_etal_1999}, echelle \citep{Ballester_1992}, PySALT \citep{Crawford_etal_2010},  and IRAF \citep{Tody_1986,Tody_1993}}.

\section*{Data availability}
The data are available as online material and can be found here: \url{https://www.dropbox.com/sh/sjye58h9sy0fyjs/AAAq8lM926pcTG2yTnE5Yydsa?dl=0}

\bibliographystyle{mnras_vanHack}
\bibliography{biblio}

\begin{thebibliography}{}
\makeatletter
\relax
\def\mn@urlcharsother{\let\do\@makeother \do\$\do\&\do\#\do\^\do\_\do\%\do\~}
\def\mn@doi{\begingroup\mn@urlcharsother \@ifnextchar [ {\mn@doi@}
  {\mn@doi@[]}}
\def\mn@doi@[#1]#2{\def\@tempa{#1}\ifx\@tempa\@empty \href
  {http://dx.doi.org/#2} {doi:#2}\else \href {http://dx.doi.org/#2} {#1}\fi
  \endgroup}
\def\mn@eprint#1#2{\mn@eprint@#1:#2::\@nil}
\def\mn@eprint@arXiv#1{\href {http://arxiv.org/abs/#1} {{\tt arXiv:#1}}}
\def\mn@eprint@dblp#1{\href {http://dblp.uni-trier.de/rec/bibtex/#1.xml}
  {dblp:#1}}
\def\mn@eprint@#1:#2:#3:#4\@nil{\def\@tempa {#1}\def\@tempb {#2}\def\@tempc
  {#3}\ifx \@tempc \@empty \let \@tempc \@tempb \let \@tempb \@tempa \fi \ifx
  \@tempb \@empty \def\@tempb {arXiv}\fi \@ifundefined
  {mn@eprint@\@tempb}{\@tempb:\@tempc}{\expandafter \expandafter \csname
  mn@eprint@\@tempb\endcsname \expandafter{\@tempc}}}

\bibitem[\protect\citeauthoryear{{Abril}, {Schmidtobreick}, {Ederoclite}  \&
  {L{\'o}pez-Sanjuan}}{{Abril} et~al.}{2020}]{Abril_etal_2020}
{Abril} J.,  {Schmidtobreick} L.,  {Ederoclite} A.,   {L{\'o}pez-Sanjuan} C.,
  2020, \mn@doi [\mnras] {10.1093/mnrasl/slz181}, \href
  {https://ui.adsabs.harvard.edu/abs/2020MNRAS.492L..40A} {492, L40}

\bibitem[\protect\citeauthoryear{{Arai}, {Isogai}, {Yamanaka}, {Akitaya}  \&
  {Uemura}}{{Arai} et~al.}{2015}]{Arai_etal_2015}
{Arai} A.,  {Isogai} M.,  {Yamanaka} M.,  {Akitaya} H.,   {Uemura} M.,  2015,
  \mn@doi [Acta Polytechnica CTU Proceedings] {10.14311/APP.2015.02.0257},
  \href {https://ui.adsabs.harvard.edu/abs/2015AcPPP...2..257A} {2, 257}

\bibitem[\protect\citeauthoryear{{Aydi} et~al.,}{{Aydi}
  et~al.}{2018}]{Aydi_etal_2018_2}
{Aydi} E.,  et~al., 2018, \mn@doi [\mnras] {10.1093/mnras/sty1759}, \href
  {http://cdsads.u-strasbg.fr/abs/2018MNRAS.480..572A} {480, 572}

\bibitem[\protect\citeauthoryear{{Aydi} et~al.,}{{Aydi}
  et~al.}{2019}]{Aydi_etal_2019_I}
{Aydi} E.,  et~al., 2019, arXiv e-prints, \href
  {https://ui.adsabs.harvard.edu/abs/2019arXiv190309232A} {}

\bibitem[\protect\citeauthoryear{{Aydi}, {Chomiuk}, {Sokolovsky}  \&
  {Steinberg}}{{Aydi} et~al.}{2020a}]{Aydi_etal_2020a}
{Aydi} E.,  {Chomiuk} L.,  {Sokolovsky} K.~V.,   {Steinberg} E.,  2020a,
  \mn@doi [Nature Astronomy] {10.1038/s41550-017-0222-1}, \href
  {http://cdsads.u-strasbg.fr/abs/2017NatAs...1..697L} {2, 697}

\bibitem[\protect\citeauthoryear{{Aydi} et~al.,}{{Aydi}
  et~al.}{2020b}]{Aydi_etal_2020b}
{Aydi} E.,  et~al., 2020b, \mn@doi [\apj] {10.3847/1538-4357/abc3bb}, \href
  {https://ui.adsabs.harvard.edu/abs/2020ApJ...905...62A} {905, 62}

\bibitem[\protect\citeauthoryear{{Aydi} et~al.,}{{Aydi}
  et~al.}{2021}]{ATel_14710}
{Aydi} E.,  et~al., 2021, The Astronomer's Telegram, \href
  {https://ui.adsabs.harvard.edu/abs/2021ATel14710....1A} {14710, 1}

\bibitem[\protect\citeauthoryear{{Aydi} et~al.,}{{Aydi}
  et~al.}{2022a}]{Aydi_etal_2022}
{Aydi} E.,  et~al., 2022a, \mn@doi [\apj] {10.3847/1538-4357/ac913b}, \href
  {https://ui.adsabs.harvard.edu/abs/2022ApJ...939....6A} {939, 6}

\bibitem[\protect\citeauthoryear{{Aydi} et~al.,}{{Aydi}
  et~al.}{2022b}]{ATel_15355}
{Aydi} E.,  et~al., 2022b, The Astronomer's Telegram, \href
  {https://ui.adsabs.harvard.edu/abs/2022ATel15355....1A} {15355, 1}

\bibitem[\protect\citeauthoryear{{Aydi} et~al.,}{{Aydi}
  et~al.}{2022c}]{ATel_15395}
{Aydi} E.,  et~al., 2022c, The Astronomer's Telegram, \href
  {https://ui.adsabs.harvard.edu/abs/2022ATel15395....1A} {15395, 1}

\bibitem[\protect\citeauthoryear{{Ballester}}{{Ballester}}{1992}]{Ballester_1992}
{Ballester} P.,  1992, in {Grosb{\o}l} P.~J.,  {de Ruijsscher} R.~C.~E.,  eds,
  European Southern Observatory Conference and Workshop Proceedings Vol. 41,
  European Southern Observatory Conference and Workshop Proceedings. p.~177

\bibitem[\protect\citeauthoryear{{Barnes} et~al.,}{{Barnes}
  et~al.}{2008}]{Barnes_etal_2008}
{Barnes} S.~I.,  et~al., 2008, in Ground-based and Airborne Instrumentation for
  Astronomy II. p. 70140K, \mn@doi{10.1117/12.788219}

\bibitem[\protect\citeauthoryear{{Belczy{\'n}ski}, {Miko{\l}ajewska}, {Munari},
  {Ivison}  \& {Friedjung}}{{Belczy{\'n}ski} et~al.}{2000}]{Belczynski_2000}
{Belczy{\'n}ski} K.,  {Miko{\l}ajewska} J.,  {Munari} U.,  {Ivison} R.~J.,
  {Friedjung} M.,  2000, \mn@doi [\aaps] {10.1051/aas:2000280}, \href
  {https://ui.adsabs.harvard.edu/abs/2000A&AS..146..407B} {146, 407}

\bibitem[\protect\citeauthoryear{{Bode} \& {Evans}}{{Bode} \&
  {Evans}}{2008}]{Bode_etal_2008}
{Bode} M.~F.,  {Evans} A.,  2008, {Classical Novae}

\bibitem[\protect\citeauthoryear{{Bode}, {Seaquist}  \& {Evans}}{{Bode}
  et~al.}{1987}]{Bode_etal_1987}
{Bode} M.~F.,  {Seaquist} E.~R.,   {Evans} A.,  1987, \mn@doi [\mnras]
  {10.1093/mnras/228.2.217}, \href
  {http://cdsads.u-strasbg.fr/abs/1987MNRAS.228..217B} {228, 217}

\bibitem[\protect\citeauthoryear{{Bonning}, {Shields}, {Stevens}  \&
  {Salviander}}{{Bonning} et~al.}{2013}]{Bonning_etal_2013}
{Bonning} E.~W.,  {Shields} G.~A.,  {Stevens} A.~C.,   {Salviander} S.,  2013,
  \mn@doi [\apj] {10.1088/0004-637X/770/1/30}, \href
  {https://ui.adsabs.harvard.edu/abs/2013ApJ...770...30B} {770, 30}

\bibitem[\protect\citeauthoryear{{Booth}, {Mohamed}  \&
  {Podsiadlowski}}{{Booth} et~al.}{2016}]{Booth_etal_2016}
{Booth} R.~A.,  {Mohamed} S.,   {Podsiadlowski} P.,  2016, \mn@doi [\mnras]
  {10.1093/mnras/stw001}, \href
  {http://cdsads.u-strasbg.fr/abs/2016MNRAS.457..822B} {457, 822}

\bibitem[\protect\citeauthoryear{{Bowen}}{{Bowen}}{1988}]{Bowen_1988}
{Bowen} G.~H.,  1988, \mn@doi [\apj] {10.1086/166378}, \href
  {http://cdsads.u-strasbg.fr/abs/1988ApJ...329..299B} {329, 299}

\bibitem[\protect\citeauthoryear{{Bramall} et~al.,}{{Bramall}
  et~al.}{2010}]{Bramall_etal_2010}
{Bramall} D.~G.,  et~al., 2010, in Ground-based and Airborne Instrumentation
  for Astronomy III. p. 77354F, \mn@doi{10.1117/12.856382}

\bibitem[\protect\citeauthoryear{{Bramall} et~al.,}{{Bramall}
  et~al.}{2012}]{Bramall_etal_2012}
{Bramall} D.~G.,  et~al., 2012, in Ground-based and Airborne Instrumentation
  for Astronomy IV. p. 84460A, \mn@doi{10.1117/12.925935}

\bibitem[\protect\citeauthoryear{{Brandi}, {Quiroga}, {Miko{\l}ajewska},
  {Ferrer}  \& {Garc{\'{\i}}a}}{{Brandi} et~al.}{2009}]{Brandi_etal_2009}
{Brandi} E.,  {Quiroga} C.,  {Miko{\l}ajewska} J.,  {Ferrer} O.~E.,
  {Garc{\'{\i}}a} L.~G.,  2009, \mn@doi [\aap] {10.1051/0004-6361/200811417},
  \href {http://cdsads.u-strasbg.fr/abs/2009A%26A...497..815B} {497, 815}

\bibitem[\protect\citeauthoryear{{Breedt} et~al.,}{{Breedt}
  et~al.}{2014}]{Breedt_etal_2014}
{Breedt} E.,  et~al., 2014, \mn@doi [\mnras] {10.1093/mnras/stu1377}, \href
  {https://ui.adsabs.harvard.edu/abs/2014MNRAS.443.3174B} {443, 3174}

\bibitem[\protect\citeauthoryear{{Brink} et~al.,}{{Brink}
  et~al.}{2022}]{ATel_15270}
{Brink} J.,  et~al., 2022, The Astronomer's Telegram, \href
  {https://ui.adsabs.harvard.edu/abs/2022ATel15270....1B} {15270, 1}

\bibitem[\protect\citeauthoryear{{Bruch}}{{Bruch}}{1989}]{Bruch_1989}
{Bruch} A.,  1989, \aaps, \href
  {https://ui.adsabs.harvard.edu/abs/1989A&AS...78..145B} {78, 145}

\bibitem[\protect\citeauthoryear{{Bruch} \& {Engel}}{{Bruch} \&
  {Engel}}{1994}]{Bruch_etal_1994}
{Bruch} A.,  {Engel} A.,  1994, \aaps, \href
  {https://ui.adsabs.harvard.edu/abs/1994A&AS..104...79B} {104, 79}

\bibitem[\protect\citeauthoryear{{Bruch} \& {Schimpke}}{{Bruch} \&
  {Schimpke}}{1992}]{Bruch_Schimpke_1992}
{Bruch} A.,  {Schimpke} T.,  1992, \aaps, \href
  {https://ui.adsabs.harvard.edu/abs/1992A&AS...93..419B} {93, 419}

\bibitem[\protect\citeauthoryear{{Buckley}, {Swart}  \& {Meiring}}{{Buckley}
  et~al.}{2006}]{Buckley_etal_2006}
{Buckley} D.~A.~H.,  {Swart} G.~P.,   {Meiring} J.~G.,  2006, in Society of
  Photo-Optical Instrumentation Engineers (SPIE) Conference Series. p. 62670Z,
  \mn@doi{10.1117/12.673750}

\bibitem[\protect\citeauthoryear{{Burgh}, {Nordsieck}, {Kobulnicky},
  {Williams}, {O'Donoghue}, {Smith}  \& {Percival}}{{Burgh}
  et~al.}{2003}]{Burgh_etal_2003}
{Burgh} E.~B.,  {Nordsieck} K.~H.,  {Kobulnicky} H.~A.,  {Williams} T.~B.,
  {O'Donoghue} D.,  {Smith} M.~P.,   {Percival} J.~W.,  2003, in {Iye} M.,
  {Moorwood} A.~F.~M.,  eds,  Society of Photo-Optical Instrumentation
  Engineers (SPIE) Conference Series Vol. 4841, Instrument Design and
  Performance for Optical/Infrared Ground-based Telescopes. pp 1463--1471,
  \mn@doi{10.1117/12.460312}

\bibitem[\protect\citeauthoryear{{Cannizzo}}{{Cannizzo}}{1993}]{Cannizzo_etal_1993}
{Cannizzo} J.~K.,  1993, \mn@doi [\apj] {10.1086/173486}, \href
  {https://ui.adsabs.harvard.edu/abs/1993ApJ...419..318C} {419, 318}

\bibitem[\protect\citeauthoryear{{Chen} et~al.,}{{Chen}
  et~al.}{2019}]{Chen_etal_2019}
{Chen} B.~Q.,  et~al., 2019, \mn@doi [\mnras] {10.1093/mnras/sty3341}, \href
  {https://ui.adsabs.harvard.edu/abs/2019MNRAS.483.4277C} {483, 4277}

\bibitem[\protect\citeauthoryear{{Chochol}, {Pribulla}, {Parimucha}  \&
  {Va{\v{n}}ko}}{{Chochol} et~al.}{2003}]{Chochol_etal_2003}
{Chochol} D.,  {Pribulla} T.,  {Parimucha} {\v{S}}.,   {Va{\v{n}}ko} M.,  2003,
  \mn@doi [Baltic Astronomy] {10.1515/astro-2017-0089}, \href
  {https://ui.adsabs.harvard.edu/abs/2003BaltA..12..610C} {12, 610}

\bibitem[\protect\citeauthoryear{{Chomiuk} et~al.,}{{Chomiuk}
  et~al.}{2014}]{Chomiuk_etal_2014}
{Chomiuk} L.,  et~al., 2014, \mn@doi [\nat] {10.1038/nature13773}, \href
  {http://cdsads.u-strasbg.fr/abs/2014Natur.514..339C} {514, 339}

\bibitem[\protect\citeauthoryear{{Chomiuk}, {Metzger}  \& {Shen}}{{Chomiuk}
  et~al.}{2021}]{Chomiuk_etal_2020}
{Chomiuk} L.,  {Metzger} B.~D.,   {Shen} K.~J.,  2021, \mn@doi [\araa]
  {10.1146/annurev-astro-112420-114502}, \href
  {https://ui.adsabs.harvard.edu/abs/2021ARA&A..59..391C} {59, 391}

\bibitem[\protect\citeauthoryear{{Clemens}, {Crain}  \& {Anderson}}{{Clemens}
  et~al.}{2004}]{Clemens_etal_2004}
{Clemens} J.~C.,  {Crain} J.~A.,   {Anderson} R.,  2004, in {Moorwood}
  A.~F.~M.,  {Iye} M.,  eds,  \procspie Vol. 5492, Ground-based Instrumentation
  for Astronomy. pp 331--340, \mn@doi{10.1117/12.550069}

\bibitem[\protect\citeauthoryear{{Crause} et~al.,}{{Crause}
  et~al.}{2014}]{Crause_etal_2014}
{Crause} L.~A.,  et~al., 2014, in Ground-based and Airborne Instrumentation for
  Astronomy V. p. 91476T, \mn@doi{10.1117/12.2055635}

\bibitem[\protect\citeauthoryear{{Crawford} et~al.,}{{Crawford}
  et~al.}{2010}]{Crawford_etal_2010}
{Crawford} S.~M.,  et~al., 2010, in Society of Photo-Optical Instrumentation
  Engineers (SPIE) Conference Series. p.~25, \mn@doi{10.1117/12.857000}

\bibitem[\protect\citeauthoryear{{Della Valle} \& {Izzo}}{{Della Valle} \&
  {Izzo}}{2020}]{Della_Valle_Izzo_2020}
{Della Valle} M.,  {Izzo} L.,  2020, \mn@doi [\aapr]
  {10.1007/s00159-020-0124-6}, \href
  {https://ui.adsabs.harvard.edu/abs/2020A&ARv..28....3D} {28, 3}

\bibitem[\protect\citeauthoryear{{Dobrzycka} \& {Kenyon}}{{Dobrzycka} \&
  {Kenyon}}{1994}]{Dobrzycka_etal_1994}
{Dobrzycka} D.,  {Kenyon} S.~J.,  1994, \mn@doi [\aj] {10.1086/117238}, \href
  {https://ui.adsabs.harvard.edu/abs/1994AJ....108.2259D} {108, 2259}

\bibitem[\protect\citeauthoryear{{Downes}, {Webbink}, {Shara}, {Ritter}, {Kolb}
   \& {Duerbeck}}{{Downes} et~al.}{2001}]{Downes_etal_2001}
{Downes} R.~A.,  {Webbink} R.~F.,  {Shara} M.~M.,  {Ritter} H.,  {Kolb} U.,
  {Duerbeck} H.~W.,  2001, \mn@doi [\pasp] {10.1086/320802}, \href
  {https://ui.adsabs.harvard.edu/abs/2001PASP..113..764D} {113, 764}

\bibitem[\protect\citeauthoryear{{Draine}}{{Draine}}{2009}]{Draine_2009}
{Draine} B.~T.,  2009, in {Henning} T.,  {Gr{\"u}n} E.,   {Steinacker} J.,
  eds,  Astronomical Society of the Pacific Conference Series Vol. 414, Cosmic
  Dust - Near and Far. p.~453 (\mn@eprint {arXiv} {0903.1658}),
  \mn@doi{10.48550/arXiv.0903.1658}

\bibitem[\protect\citeauthoryear{{Drew} et~al.,}{{Drew}
  et~al.}{2014}]{Drew_etal_2014}
{Drew} J.~E.,  et~al., 2014, \mn@doi [\mnras] {10.1093/mnras/stu394}, \href
  {https://ui.adsabs.harvard.edu/abs/2014MNRAS.440.2036D} {440, 2036}

\bibitem[\protect\citeauthoryear{{Ederoclite}}{{Ederoclite}}{2014}]{Ederoclite_2014}
{Ederoclite} A.,  2014, in {Woudt} P.~A.,  {Ribeiro} V.~A.~R.~M.,  eds,
  Astronomical Society of the Pacific Conference Series Vol. 490, Stellar
  Novae: Past and Future Decades. p.~163 (\mn@eprint {arXiv} {1304.1305})

\bibitem[\protect\citeauthoryear{{Friedjung}}{{Friedjung}}{1992}]{Friedjung_1992}
{Friedjung} M.,  1992, \aap, \href
  {http://cdsads.u-strasbg.fr/abs/1992A%26A...262..487F} {262, 487}

\bibitem[\protect\citeauthoryear{{Friedman} et~al.,}{{Friedman}
  et~al.}{2011}]{Friedman_etal_2011}
{Friedman} S.~D.,  et~al., 2011, \mn@doi [\apj] {10.1088/0004-637X/727/1/33},
  \href {https://ui.adsabs.harvard.edu/abs/2011ApJ...727...33F} {727, 33}

\bibitem[\protect\citeauthoryear{{Gaia Collaboration} et~al.,}{{Gaia
  Collaboration} et~al.}{2021}]{2021A&A...649A...1G}
{Gaia Collaboration} et~al., 2021, \mn@doi [\aap]
  {10.1051/0004-6361/202039657}, \href
  {https://ui.adsabs.harvard.edu/abs/2021A&A...649A...1G} {649, A1}

\bibitem[\protect\citeauthoryear{{Gaia Collaboration} et~al.,}{{Gaia
  Collaboration} et~al.}{2022}]{2022arXiv220800211G}
{Gaia Collaboration} et~al., 2022, arXiv e-prints, \href
  {https://ui.adsabs.harvard.edu/abs/2022arXiv220800211G} {p. arXiv:2208.00211}

\bibitem[\protect\citeauthoryear{{Gaposchkin}}{{Gaposchkin}}{1957}]{Payne-Gaposchkin_1957}
{Gaposchkin} C.~H.~P.,  1957, {The galactic novae.}

\bibitem[\protect\citeauthoryear{{Godon}, {Sion}, {Williams}  \&
  {Starrfield}}{{Godon} et~al.}{2018}]{Godon_etal_2018}
{Godon} P.,  {Sion} E.~M.,  {Williams} R.~E.,   {Starrfield} S.,  2018, \mn@doi
  [\apj] {10.3847/1538-4357/aacd0a}, \href
  {https://ui.adsabs.harvard.edu/abs/2018ApJ...862...89G} {862, 89}

\bibitem[\protect\citeauthoryear{{Goranskij} et~al.,}{{Goranskij}
  et~al.}{2007}]{Goranskij_etal_2007}
{Goranskij} V.~P.,  et~al., 2007, \mn@doi [Astrophysical Bulletin]
  {10.1134/S1990341307020046}, \href
  {http://cdsads.u-strasbg.fr/abs/2007AstBu..62..125G} {62, 125}

\bibitem[\protect\citeauthoryear{{Hachisu} \& {Kato}}{{Hachisu} \&
  {Kato}}{2004}]{Hachisu_Kato_2004}
{Hachisu} I.,  {Kato} M.,  2004, \mn@doi [\apjl] {10.1086/424595}, \href
  {http://cdsads.u-strasbg.fr/abs/2004ApJ...612L..57H} {612, L57}

\bibitem[\protect\citeauthoryear{{Hachisu} \& {Kato}}{{Hachisu} \&
  {Kato}}{2022}]{Hachisu_Kato_2022}
{Hachisu} I.,  {Kato} M.,  2022, \mn@doi [\apj] {10.3847/1538-4357/ac9475},
  \href {https://ui.adsabs.harvard.edu/abs/2022ApJ...939....1H} {939, 1}

\bibitem[\protect\citeauthoryear{{Han}, {Boonrucksar}, {Qian}, {Xiaohui},
  {Wang}, {Zhu}, {Dong}  \& {Zhi}}{{Han} et~al.}{2020}]{Han_etal_2020}
{Han} Z.,  {Boonrucksar} S.,  {Qian} S.,  {Xiaohui} F.,  {Wang} Q.,  {Zhu} L.,
  {Dong} A.,   {Zhi} Q.,  2020, \mn@doi [\pasj] {10.1093/pasj/psaa065}, \href
  {https://ui.adsabs.harvard.edu/abs/2020PASJ...72...76H} {72, 76}

\bibitem[\protect\citeauthoryear{{Harrison}, {Johnson}  \&
  {Spyromilio}}{{Harrison} et~al.}{1993}]{Harrison_etal_1993}
{Harrison} T.~E.,  {Johnson} J.~J.,   {Spyromilio} J.,  1993, \mn@doi [\aj]
  {10.1086/116429}, \href
  {https://ui.adsabs.harvard.edu/abs/1993AJ....105..320H} {105, 320}

\bibitem[\protect\citeauthoryear{{Harvey}, {Redman}, {Darnley}, {Williams},
  {Berdyugin}, {Piirola}, {Fitzgerald}  \& {O'Connor}}{{Harvey}
  et~al.}{2018}]{Harvey2018}
{Harvey} E.~J.,  {Redman} M.~P.,  {Darnley} M.~J.,  {Williams} S.~C.,
  {Berdyugin} A.,  {Piirola} V.~E.,  {Fitzgerald} K.~P.,   {O'Connor} E.~G.~P.,
   2018, \mn@doi [\aap] {10.1051/0004-6361/201731741}, \href
  {https://ui.adsabs.harvard.edu/abs/2018A&A...611A...3H} {611, A3}

\bibitem[\protect\citeauthoryear{{Hillman}, {Prialnik}, {Kovetz}, {Shara}  \&
  {Neill}}{{Hillman} et~al.}{2014}]{Hillman_etal_2014}
{Hillman} Y.,  {Prialnik} D.,  {Kovetz} A.,  {Shara} M.~M.,   {Neill} J.~D.,
  2014, \mn@doi [\mnras] {10.1093/mnras/stt2027}, \href
  {http://adsabs.harvard.edu/abs/2014MNRAS.437.1962H} {437, 1962}

\bibitem[\protect\citeauthoryear{{Hodgkin} et~al.,}{{Hodgkin}
  et~al.}{2022}]{2022TNSTR.313....1H}
{Hodgkin} S.~T.,  et~al., 2022, Transient Name Server Discovery Report, \href
  {https://ui.adsabs.harvard.edu/abs/2022TNSTR.313....1H} {2022-313, 1}

\bibitem[\protect\citeauthoryear{{Hou}, {Luo}, {Li}  \& {Qin}}{{Hou}
  et~al.}{2020}]{Hou_etal_2020}
{Hou} W.,  {Luo} A.~l.,  {Li} Y.-B.,   {Qin} L.,  2020, \mn@doi [\aj]
  {10.3847/1538-3881/ab5962}, \href
  {https://ui.adsabs.harvard.edu/abs/2020AJ....159...43H} {159, 43}

\bibitem[\protect\citeauthoryear{{Hou}, {Luo}, {Dong}, {Chen}  \& {Bai}}{{Hou}
  et~al.}{2023}]{Hou_etal_2023}
{Hou} W.,  {Luo} A.~L.,  {Dong} Y.-Q.,  {Chen} X.-L.,   {Bai} Z.-R.,  2023,
  \mn@doi [\aj] {10.3847/1538-3881/aca906}, \href
  {https://ui.adsabs.harvard.edu/abs/2023AJ....165..148H} {165, 148}

\bibitem[\protect\citeauthoryear{{Hounsell} et~al.,}{{Hounsell}
  et~al.}{2010}]{Hounsell_etal_2010}
{Hounsell} R.,  et~al., 2010, \mn@doi [\apj] {10.1088/0004-637X/724/1/480},
  \href {http://cdsads.u-strasbg.fr/abs/2010ApJ...724..480H} {724, 480}

\bibitem[\protect\citeauthoryear{{Hounsell} et~al.,}{{Hounsell}
  et~al.}{2016}]{Hounsell_etal_2016}
{Hounsell} R.,  et~al., 2016, \mn@doi [\apj] {10.3847/0004-637X/820/2/104},
  \href {https://ui.adsabs.harvard.edu/abs/2016ApJ...820..104H} {820, 104}

\bibitem[\protect\citeauthoryear{{Izzo} et~al.,}{{Izzo}
  et~al.}{2012}]{Izzo_etal_2012}
{Izzo} L.,  et~al., 2012, \memsai, \href
  {https://ui.adsabs.harvard.edu/abs/2012MmSAI..83..830I} {83, 830}

\bibitem[\protect\citeauthoryear{{Jiang}, {Luo}, {Zhao}  \& {Wei}}{{Jiang}
  et~al.}{2013}]{Jiang_etal_2013}
{Jiang} B.,  {Luo} A.,  {Zhao} Y.,   {Wei} P.,  2013, \mn@doi [\mnras]
  {10.1093/mnras/sts665}, \href
  {https://ui.adsabs.harvard.edu/abs/2013MNRAS.430..986J} {430, 986}

\bibitem[\protect\citeauthoryear{{Kafka}}{{Kafka}}{2020}]{Kafka_2020}
{Kafka} S.,  2020, Observations from the AAVSO International Database, \href
  {https://ui.adsabs.harvard.edu/abs/2020ATel13731....1P} {\url{
  https://www.aavso.org}}

\bibitem[\protect\citeauthoryear{{Kato} \& {Takamizawa}}{{Kato} \&
  {Takamizawa}}{2001}]{Kato_Takamizawa_2001}
{Kato} T.,  {Takamizawa} K.,  2001, Information Bulletin on Variable Stars,
  \href {http://cdsads.u-strasbg.fr/abs/2001IBVS.5100....1K} {5100}

\bibitem[\protect\citeauthoryear{{Kato}, {Maehara}  \& {Uemura}}{{Kato}
  et~al.}{2012}]{Kato_etal_2012}
{Kato} T.,  {Maehara} H.,   {Uemura} M.,  2012, \mn@doi [\pasj]
  {10.1093/pasj/64.3.63}, \href
  {https://ui.adsabs.harvard.edu/abs/2012PASJ...64...63K} {64, 63}

\bibitem[\protect\citeauthoryear{{Kawash} et~al.,}{{Kawash}
  et~al.}{2021a}]{2021ApJ...910..120K}
{Kawash} A.,  et~al., 2021a, \mn@doi [\apj] {10.3847/1538-4357/abe53d}, \href
  {https://ui.adsabs.harvard.edu/abs/2021ApJ...910..120K} {910, 120}

\bibitem[\protect\citeauthoryear{{Kawash} et~al.,}{{Kawash}
  et~al.}{2021b}]{Kawash_etal_2021a}
{Kawash} A.,  et~al., 2021b, \mn@doi [\apj] {10.3847/1538-4357/abe53d}, \href
  {https://ui.adsabs.harvard.edu/abs/2021ApJ...910..120K} {910, 120}

\bibitem[\protect\citeauthoryear{{Kenyon} \& {Truran}}{{Kenyon} \&
  {Truran}}{1983}]{Kenyon_Turan_1983}
{Kenyon} S.~J.,  {Truran} J.~W.,  1983, \mn@doi [\apj] {10.1086/161367}, \href
  {https://ui.adsabs.harvard.edu/abs/1983ApJ...273..280K} {273, 280}

\bibitem[\protect\citeauthoryear{{Kniazev}, {Gvaramadze}  \&
  {Berdnikov}}{{Kniazev} et~al.}{2016}]{kniazev_etal_2016}
{Kniazev} A.~Y.,  {Gvaramadze} V.~V.,   {Berdnikov} L.~N.,  2016, \mn@doi
  [\mnras] {10.1093/mnras/stw889}, \href
  {https://ui.adsabs.harvard.edu/abs/2016MNRAS.459.3068K} {459, 3068}

\bibitem[\protect\citeauthoryear{{Kobulnicky}, {Nordsieck}, {Burgh}, {Smith},
  {Percival}, {Williams}  \& {O'Donoghue}}{{Kobulnicky}
  et~al.}{2003}]{Kobulnicky_etal_2003}
{Kobulnicky} H.~A.,  {Nordsieck} K.~H.,  {Burgh} E.~B.,  {Smith} M.~P.,
  {Percival} J.~W.,  {Williams} T.~B.,   {O'Donoghue} D.,  2003, in {Iye} M.,
  {Moorwood} A.~F.~M.,  eds,  Society of Photo-Optical Instrumentation
  Engineers (SPIE) Conference Series Vol. 4841, Instrument Design and
  Performance for Optical/Infrared Ground-based Telescopes. pp 1634--1644,
  \mn@doi{10.1117/12.460315}

\bibitem[\protect\citeauthoryear{{Kochanek} et~al.,}{{Kochanek}
  et~al.}{2017}]{Kochanek_etal_2017}
{Kochanek} C.~S.,  et~al., 2017, \mn@doi [\pasp] {10.1088/1538-3873/aa80d9},
  \href {http://adsabs.harvard.edu/abs/2017PASP..129j4502K} {129, 104502}

\bibitem[\protect\citeauthoryear{{K{\"o}nig} et~al.,}{{K{\"o}nig}
  et~al.}{2022}]{Konig_etal_2022}
{K{\"o}nig} O.,  et~al., 2022, \mn@doi [\nat] {10.1038/s41586-022-04635-y},
  \href {https://ui.adsabs.harvard.edu/abs/2022Natur.605..248K} {605, 248}

\bibitem[\protect\citeauthoryear{{Krautter}}{{Krautter}}{2002}]{Krautter_2002}
{Krautter} J.,  2002, in {Hernanz} M.,  {Jos{\'e}} J.,  eds,  American
  Institute of Physics Conference Series Vol. 637, Classical Nova Explosions.
  pp 345--354, \mn@doi{10.1063/1.1518228}

\bibitem[\protect\citeauthoryear{{Luna} et~al.,}{{Luna}
  et~al.}{2018}]{Luna_etal_2018}
{Luna} G.~J.~M.,  et~al., 2018, \mn@doi [\aap] {10.1051/0004-6361/201833747},
  \href {https://ui.adsabs.harvard.edu/abs/2018A&A...619A..61L} {619, A61}

\bibitem[\protect\citeauthoryear{{Marshall}, {Robin}, {Reyl{\'e}}, {Schultheis}
   \& {Picaud}}{{Marshall} et~al.}{2006}]{Marshall_etal_2006}
{Marshall} D.~J.,  {Robin} A.~C.,  {Reyl{\'e}} C.,  {Schultheis} M.,   {Picaud}
  S.,  2006, \mn@doi [\aap] {10.1051/0004-6361:20053842}, \href
  {https://ui.adsabs.harvard.edu/abs/2006A&A...453..635M} {453, 635}

\bibitem[\protect\citeauthoryear{{Matsumoto} \& {Metzger}}{{Matsumoto} \&
  {Metzger}}{2022}]{Matsumoto_Metzger_2022}
{Matsumoto} T.,  {Metzger} B.~D.,  2022, \mn@doi [\apj]
  {10.3847/1538-4357/ac6269}, \href
  {https://ui.adsabs.harvard.edu/abs/2022ApJ...938....5M} {938, 5}

\bibitem[\protect\citeauthoryear{{McLaughlin}}{{McLaughlin}}{1944}]{McLaughlin_1944}
{McLaughlin} D.~B.,  1944, Popular Astronomy, \href
  {http://adsabs.harvard.edu/abs/1944PA.....52..109M} {52, 109}

\bibitem[\protect\citeauthoryear{{Mclaughlin}}{{Mclaughlin}}{1947}]{McLaughlin_1947}
{Mclaughlin} D.~B.,  1947, \mn@doi [\pasp] {10.1086/125957}, \href
  {https://ui.adsabs.harvard.edu/abs/1947PASP...59..244M} {59, 244}

\bibitem[\protect\citeauthoryear{{Miko{\l}ajewska}}{{Miko{\l}ajewska}}{2003}]{Mikolajewska_2003}
{Miko{\l}ajewska} J.,  2003, in {Corradi} R.~L.~M.,  {Mikolajewska} J.,
  {Mahoney} T.~J.,  eds,  Astronomical Society of the Pacific Conference Series
  Vol. 303, Symbiotic Stars Probing Stellar Evolution. p.~9 (\mn@eprint {arXiv}
  {astro-ph/0210489}), \mn@doi{10.48550/arXiv.astro-ph/0210489}

\bibitem[\protect\citeauthoryear{{Miko{\l}ajewska}}{{Miko{\l}ajewska}}{2007}]{Mikolajewska_2007}
{Miko{\l}ajewska} J.,  2007, Baltic Astronomy, \href
  {https://ui.adsabs.harvard.edu/abs/2007BaltA..16....1M} {16, 1}

\bibitem[\protect\citeauthoryear{{Mikolajewska}}{{Mikolajewska}}{2010}]{Mikolajewska_2010}
{Mikolajewska} J.,  2010, \mn@doi [arXiv e-prints] {10.48550/arXiv.1011.5657},
  \href {https://ui.adsabs.harvard.edu/abs/2010arXiv1011.5657M} {p.
  arXiv:1011.5657}

\bibitem[\protect\citeauthoryear{{Miko{\l}ajewska}, {I{\l}kiewicz}, {Ga{\l}an},
  {Monard}, {Otulakowska-Hypka}, {Shara}  \& {Udalski}}{{Miko{\l}ajewska}
  et~al.}{2021}]{Mikoljewska_etal_2021}
{Miko{\l}ajewska} J.,  {I{\l}kiewicz} K.,  {Ga{\l}an} C.,  {Monard} B.,
  {Otulakowska-Hypka} M.,  {Shara} M.~M.,   {Udalski} A.,  2021, \mn@doi
  [\mnras] {10.1093/mnras/stab1058}, \href
  {https://ui.adsabs.harvard.edu/abs/2021MNRAS.504.2122M} {504, 2122}

\bibitem[\protect\citeauthoryear{{Mondal}, {Anupama}, {Kamath}, {Das},
  {Selvakumar}  \& {Mondal}}{{Mondal} et~al.}{2018}]{Mondal_etal_2018}
{Mondal} A.,  {Anupama} G.~C.,  {Kamath} U.~S.,  {Das} R.,  {Selvakumar} G.,
  {Mondal} S.,  2018, \mn@doi [\mnras] {10.1093/mnras/stx2988}, \href
  {https://ui.adsabs.harvard.edu/abs/2018MNRAS.474.4211M} {474, 4211}

\bibitem[\protect\citeauthoryear{{Morales-Rueda} \& {Marsh}}{{Morales-Rueda} \&
  {Marsh}}{2002}]{Morales-Rueda_Marsh_2002}
{Morales-Rueda} L.,  {Marsh} T.~R.,  2002, \mn@doi [\mnras]
  {10.1046/j.1365-8711.2002.05357.x}, \href
  {https://ui.adsabs.harvard.edu/abs/2002MNRAS.332..814M} {332, 814}

\bibitem[\protect\citeauthoryear{{Morisset} \& {Pequignot}}{{Morisset} \&
  {Pequignot}}{1996}]{Morisset_Pequignot_1996}
{Morisset} C.,  {Pequignot} D.,  1996, \aap, \href
  {https://ui.adsabs.harvard.edu/abs/1996A&A...312..135M} {312, 135}

\bibitem[\protect\citeauthoryear{{Mr{\'o}z} et~al.,}{{Mr{\'o}z}
  et~al.}{2016}]{mroz16}
{Mr{\'o}z} P.,  et~al., 2016, \mn@doi [\nat] {10.1038/nature19066}, \href
  {https://ui.adsabs.harvard.edu/abs/2016Natur.537..649M} {537, 649}

\bibitem[\protect\citeauthoryear{{Munari} \& {Zwitter}}{{Munari} \&
  {Zwitter}}{1997}]{Munari_Zwitter_1997}
{Munari} U.,  {Zwitter} T.,  1997, \aap, \href
  {http://cdsads.u-strasbg.fr/abs/1997A%26A...318..269M} {318, 269}

\bibitem[\protect\citeauthoryear{{Munari} et~al.,}{{Munari}
  et~al.}{1996}]{Munari_etal_1996}
{Munari} U.,  et~al., 1996, \aap, \href
  {http://cdsads.u-strasbg.fr/abs/1996A%26A...315..166M} {315, 166}

\bibitem[\protect\citeauthoryear{{Munari}, {Ochner}, {Siviero}, {Jones},
  {Moretti}, {Tomaselli}  \& {Dallaporta}}{{Munari}
  et~al.}{2009}]{Munari_etal_2009}
{Munari} U.,  {Ochner} P.,  {Siviero} A.,  {Jones} A.,  {Moretti} S.,
  {Tomaselli} S.,   {Dallaporta} S.,  2009, Baltic Astronomy, \href
  {https://ui.adsabs.harvard.edu/abs/2009BaltA..18...75M} {18, 75}

\bibitem[\protect\citeauthoryear{{Munari}, {Ribeiro}, {Bode}  \&
  {Saguner}}{{Munari} et~al.}{2011a}]{Munari_etal_2011}
{Munari} U.,  {Ribeiro} V.~A.~R.~M.,  {Bode} M.~F.,   {Saguner} T.,  2011a,
  \mn@doi [\mnras] {10.1111/j.1365-2966.2010.17462.x}, \href
  {http://cdsads.u-strasbg.fr/abs/2011MNRAS.410..525M} {410, 525}

\bibitem[\protect\citeauthoryear{{Munari} et~al.,}{{Munari}
  et~al.}{2011b}]{Munari_etal_2011_V407cyg}
{Munari} U.,  et~al., 2011b, \mn@doi [\mnras]
  {10.1111/j.1745-3933.2010.00979.x}, \href
  {https://ui.adsabs.harvard.edu/abs/2011MNRAS.410L..52M} {410, L52}

\bibitem[\protect\citeauthoryear{{Munari}, {Dallaporta}  \& {Cherini}}{{Munari}
  et~al.}{2016}]{Munari_etal_2016}
{Munari} U.,  {Dallaporta} S.,   {Cherini} G.,  2016, \mn@doi [\na]
  {10.1016/j.newast.2016.01.002}, \href
  {https://ui.adsabs.harvard.edu/abs/2016NewA...47....7M} {47, 7}

\bibitem[\protect\citeauthoryear{{Murphy-Glaysher} et~al.,}{{Murphy-Glaysher}
  et~al.}{2022}]{Murphy-Glaysher_etal_2022}
{Murphy-Glaysher} F.~J.,  et~al., 2022, \mn@doi [\mnras]
  {10.1093/mnras/stac1577}, \href
  {https://ui.adsabs.harvard.edu/abs/2022MNRAS.514.6183M} {514, 6183}

\bibitem[\protect\citeauthoryear{{O'Brien} et~al.,}{{O'Brien}
  et~al.}{2006}]{Obrien_etal_2006}
{O'Brien} T.~J.,  et~al., 2006, \mn@doi [\nat] {10.1038/nature04949}, \href
  {https://ui.adsabs.harvard.edu/abs/2006Natur.442..279O} {442, 279}

\bibitem[\protect\citeauthoryear{{O'Donoghue} et~al.,}{{O'Donoghue}
  et~al.}{2006}]{Odonoghue_etal_2006}
{O'Donoghue} D.,  et~al., 2006, \mn@doi [\mnras]
  {10.1111/j.1365-2966.2006.10834.x}, \href
  {http://cdsads.u-strasbg.fr/abs/2006MNRAS.372..151O} {372, 151}

\bibitem[\protect\citeauthoryear{{Orio} et~al.,}{{Orio}
  et~al.}{2020}]{Orio_etal_2020}
{Orio} M.,  et~al., 2020, \mn@doi [\apj] {10.3847/1538-4357/ab8c4d}, \href
  {https://ui.adsabs.harvard.edu/abs/2020ApJ...895...80O} {895, 80}

\bibitem[\protect\citeauthoryear{{Pastorello} et~al.,}{{Pastorello}
  et~al.}{2019}]{Pastorello_etal_2019}
{Pastorello} A.,  et~al., 2019, \mn@doi [\aap] {10.1051/0004-6361/201935999},
  \href {https://ui.adsabs.harvard.edu/abs/2019A&A...630A..75P} {630, A75}

\bibitem[\protect\citeauthoryear{{Patterson} et~al.,}{{Patterson}
  et~al.}{1998}]{Patterson_etal_1998}
{Patterson} J.,  et~al., 1998, \mn@doi [\pasp] {10.1086/316147}, \href
  {https://ui.adsabs.harvard.edu/abs/1998PASP..110..380P} {110, 380}

\bibitem[\protect\citeauthoryear{{Pejcha}}{{Pejcha}}{2009}]{Pejcha_2009}
{Pejcha} O.,  2009, \mn@doi [\apjl] {10.1088/0004-637X/701/2/L119}, \href
  {http://cdsads.u-strasbg.fr/abs/2009ApJ...701L.119P} {701, L119}

\bibitem[\protect\citeauthoryear{{Pejcha}, {Metzger}  \& {Tomida}}{{Pejcha}
  et~al.}{2016}]{Pejcha_etal_2016}
{Pejcha} O.,  {Metzger} B.~D.,   {Tomida} K.,  2016, \mn@doi [\mnras]
  {10.1093/mnras/stw1481}, \href
  {https://ui.adsabs.harvard.edu/abs/2016MNRAS.461.2527P} {461, 2527}

\bibitem[\protect\citeauthoryear{{Pejcha}, {Metzger}, {Tyles}  \&
  {Tomida}}{{Pejcha} et~al.}{2017}]{Pejcha_etal_2017}
{Pejcha} O.,  {Metzger} B.~D.,  {Tyles} J.~G.,   {Tomida} K.,  2017, \mn@doi
  [\apj] {10.3847/1538-4357/aa95b9}, \href
  {https://ui.adsabs.harvard.edu/abs/2017ApJ...850...59P} {850, 59}

\bibitem[\protect\citeauthoryear{{Prialnik}}{{Prialnik}}{1986}]{Prialnik_1986}
{Prialnik} D.,  1986, \mn@doi [\apj] {10.1086/164677}, \href
  {http://adsabs.harvard.edu/abs/1986ApJ...310..222P} {310, 222}

\bibitem[\protect\citeauthoryear{{Salazar}, {LeBleu}, {Schaefer}, {Landolt}  \&
  {Dvorak}}{{Salazar} et~al.}{2017}]{Salazar_etal_2017}
{Salazar} I.~V.,  {LeBleu} A.,  {Schaefer} B.~E.,  {Landolt} A.~U.,   {Dvorak}
  S.,  2017, \mn@doi [\mnras] {10.1093/mnras/stx1161}, \href
  {https://ui.adsabs.harvard.edu/abs/2017MNRAS.469.4116V} {469, 4116}

\bibitem[\protect\citeauthoryear{{Schaefer}}{{Schaefer}}{2022}]{2022MNRAS.517.6150S}
{Schaefer} B.~E.,  2022, \mn@doi [\mnras] {10.1093/mnras/stac2900}, \href
  {https://ui.adsabs.harvard.edu/abs/2022MNRAS.517.6150S} {517, 6150}

\bibitem[\protect\citeauthoryear{{Schaefer} \& {Collazzi}}{{Schaefer} \&
  {Collazzi}}{2010}]{Schaefer_Collazi_2010}
{Schaefer} B.~E.,  {Collazzi} A.~C.,  2010, \mn@doi [\aj]
  {10.1088/0004-6256/139/5/1831}, \href
  {https://ui.adsabs.harvard.edu/abs/2010AJ....139.1831S} {139, 1831}

\bibitem[\protect\citeauthoryear{{Schaefer} et~al.,}{{Schaefer}
  et~al.}{2013}]{Schaefer_etal_2013}
{Schaefer} B.~E.,  et~al., 2013, \mn@doi [\apj] {10.1088/0004-637X/773/1/55},
  \href {https://ui.adsabs.harvard.edu/abs/2013ApJ...773...55S} {773, 55}

\bibitem[\protect\citeauthoryear{{Schlafly} \& {Finkbeiner}}{{Schlafly} \&
  {Finkbeiner}}{2011}]{Schlafly_etal_2011}
{Schlafly} E.~F.,  {Finkbeiner} D.~P.,  2011, \mn@doi [\apj]
  {10.1088/0004-637X/737/2/103}, \href
  {http://adsabs.harvard.edu/abs/2011ApJ...737..103S} {737, 103}

\bibitem[\protect\citeauthoryear{{Schlafly} et~al.,}{{Schlafly}
  et~al.}{2018}]{Schlafly_etal_2018}
{Schlafly} E.~F.,  et~al., 2018, \mn@doi [\apjs] {10.3847/1538-4365/aaa3e2},
  \href {https://ui.adsabs.harvard.edu/abs/2018ApJS..234...39S} {234, 39}

\bibitem[\protect\citeauthoryear{{Shafter}}{{Shafter}}{2017a}]{2017ApJ...834..196S}
{Shafter} A.~W.,  2017a, \mn@doi [\apj] {10.3847/1538-4357/834/2/196}, \href
  {https://ui.adsabs.harvard.edu/abs/2017ApJ...834..196S} {834, 196}

\bibitem[\protect\citeauthoryear{{Shafter}}{{Shafter}}{2017b}]{Shafter_2017}
{Shafter} A.~W.,  2017b, \mn@doi [\apj] {10.3847/1538-4357/834/2/196}, \href
  {https://ui.adsabs.harvard.edu/abs/2017ApJ...834..196S} {834, 196}

\bibitem[\protect\citeauthoryear{{Shappee} et~al.,}{{Shappee}
  et~al.}{2014}]{Shappee_etal_2014}
{Shappee} B.~J.,  et~al., 2014, \mn@doi [\apj] {10.1088/0004-637X/788/1/48},
  \href {http://adsabs.harvard.edu/abs/2014ApJ...788...48S} {788, 48}

\bibitem[\protect\citeauthoryear{{Shara} et~al.,}{{Shara}
  et~al.}{2017}]{Shara_etal_2017_apr}
{Shara} M.~M.,  et~al., 2017, \mn@doi [\apj] {10.3847/1538-4357/aa65cd}, \href
  {http://cdsads.u-strasbg.fr/abs/2017ApJ...839..109S} {839, 109}

\bibitem[\protect\citeauthoryear{{Shen} \& {Quataert}}{{Shen} \&
  {Quataert}}{2022}]{Shen_Quataert_2022}
{Shen} K.~J.,  {Quataert} E.,  2022, \mn@doi [\apj] {10.3847/1538-4357/ac9136},
  \href {https://ui.adsabs.harvard.edu/abs/2022ApJ...938...31S} {938, 31}

\bibitem[\protect\citeauthoryear{{Shore}}{{Shore}}{2014}]{Shore_2014}
{Shore} S.~N.,  2014, in {Woudt} P.~A.,  {Ribeiro} V.~A.~R.~M.,  eds,
  Astronomical Society of the Pacific Conference Series Vol. 490, Stellar
  Novae: Past and Future Decades. p.~145

\bibitem[\protect\citeauthoryear{{Shore}, {Kuin}, {Mason}  \& {De Gennaro
  Aquino}}{{Shore} et~al.}{2018}]{Shore_etal_2018}
{Shore} S.~N.,  {Kuin} N.~P.,  {Mason} E.,   {De Gennaro Aquino} I.,  2018,
  \mn@doi [\aap] {10.1051/0004-6361/201833204}, \href
  {http://cdsads.u-strasbg.fr/abs/2018A%26A...619A.104S} {619, A104}

\bibitem[\protect\citeauthoryear{{Shu}, {Lubow}  \& {Anderson}}{{Shu}
  et~al.}{1979}]{Shu_etal_1979}
{Shu} F.~H.,  {Lubow} S.~H.,   {Anderson} L.,  1979, \mn@doi [\apj]
  {10.1086/156948}, \href
  {https://ui.adsabs.harvard.edu/abs/1979ApJ...229..223S} {229, 223}

\bibitem[\protect\citeauthoryear{{Smale} et~al.,}{{Smale}
  et~al.}{1988}]{Smale_etal_1988}
{Smale} A.~P.,  et~al., 1988, \mn@doi [\mnras] {10.1093/mnras/233.1.51}, \href
  {https://ui.adsabs.harvard.edu/abs/1988MNRAS.233...51S} {233, 51}

\bibitem[\protect\citeauthoryear{{Stahl}, {Kaufer}  \& {Tubbesing}}{{Stahl}
  et~al.}{1999}]{Stahl_etal_1999}
{Stahl} O.,  {Kaufer} A.,   {Tubbesing} S.,  1999, in {Guenther} E.,
  {Stecklum} B.,   {Klose} S.,  eds,  Astronomical Society of the Pacific
  Conference Series Vol. 188, Optical and Infrared Spectroscopy of
  Circumstellar Matter. p.~331

\bibitem[\protect\citeauthoryear{{Starrfield}}{{Starrfield}}{1989}]{Starrfield_1989_bode}
{Starrfield} S.,  1989, {Classical novae}

\bibitem[\protect\citeauthoryear{{Starrfield}, {Sparks}  \&
  {Truran}}{{Starrfield} et~al.}{1974}]{Starrfield_etal_1974}
{Starrfield} S.,  {Sparks} W.~M.,   {Truran} J.~W.,  1974, \mn@doi [\apjs]
  {10.1086/190317}, \href {http://adsabs.harvard.edu/abs/1974ApJS...28..247S}
  {28, 247}

\bibitem[\protect\citeauthoryear{{Starrfield}, {Truran}, {Sparks}, {Krautter}
  \& {MacDonald}}{{Starrfield} et~al.}{1990}]{Starrfield_etal_1990}
{Starrfield} S.,  {Truran} J.~W.,  {Sparks} W.~M.,  {Krautter} J.,
  {MacDonald} J.,  1990, in {Cassatella} A.,  {Viotti} R.,  eds, , Vol.~369,
  IAU Colloq. 122: Physics of Classical Novae.
p.~306, \mn@doi{10.1007/3-540-53500-4_143}

\bibitem[\protect\citeauthoryear{{Strope}, {Schaefer}  \& {Henden}}{{Strope}
  et~al.}{2010}]{Strope_etal_2010}
{Strope} R.~J.,  {Schaefer} B.~E.,   {Henden} A.~A.,  2010, \mn@doi [\aj]
  {10.1088/0004-6256/140/1/34}, \href
  {http://adsabs.harvard.edu/abs/2010AJ....140...34S} {140, 34}

\bibitem[\protect\citeauthoryear{{Szkody}}{{Szkody}}{1994}]{Szkod_1994}
{Szkody} P.,  1994, \mn@doi [\aj] {10.1086/117098}, \href
  {https://ui.adsabs.harvard.edu/abs/1994AJ....108..639S} {108, 639}

\bibitem[\protect\citeauthoryear{{Szkody} et~al.,}{{Szkody}
  et~al.}{2011}]{Szkody_etal_2011}
{Szkody} P.,  et~al., 2011, \mn@doi [\aj] {10.1088/0004-6256/142/6/181}, \href
  {https://ui.adsabs.harvard.edu/abs/2011AJ....142..181S} {142, 181}

\bibitem[\protect\citeauthoryear{{Tanaka}, {Nogami}, {Fujii}, {Ayani}  \&
  {Kato}}{{Tanaka} et~al.}{2011a}]{Tanaka_etal_2011_159}
{Tanaka} J.,  {Nogami} D.,  {Fujii} M.,  {Ayani} K.,   {Kato} T.,  2011a,
  \mn@doi [\pasj] {10.1093/pasj/63.1.159}, \href
  {http://cdsads.u-strasbg.fr/abs/2011PASJ...63..159T} {63, 159}

\bibitem[\protect\citeauthoryear{{Tanaka}, {Nogami}, {Fujii}, {Ayani}, {Kato},
  {Maehara}, {Kiyota}  \& {Nakajima}}{{Tanaka}
  et~al.}{2011b}]{Tanaka_etal_2011}
{Tanaka} J.,  {Nogami} D.,  {Fujii} M.,  {Ayani} K.,  {Kato} T.,  {Maehara} H.,
   {Kiyota} S.,   {Nakajima} K.,  2011b, \mn@doi [\pasj]
  {10.1093/pasj/63.4.911}, \href
  {http://adsabs.harvard.edu/abs/2011PASJ...63..911T} {63, 911}

\bibitem[\protect\citeauthoryear{{Tody}}{{Tody}}{1986}]{Tody_1986}
{Tody} D.,  1986, in {Crawford} D.~L.,  ed.,  Society of Photo-Optical
  Instrumentation Engineers (SPIE) Conference Series Vol. 627, Instrumentation
  in astronomy VI. p.~733

\bibitem[\protect\citeauthoryear{{Tody}}{{Tody}}{1993}]{Tody_1993}
{Tody} D.,  1993, in {Hanisch} R.~J.,  {Brissenden} R.~J.~V.,   {Barnes} J.,
  eds,  Astronomical Society of the Pacific Conference Series Vol. 52,
  Astronomical Data Analysis Software and Systems II. p.~173

\bibitem[\protect\citeauthoryear{{Udalski}, {Szyma{\'n}ski}  \&
  {Szyma{\'n}ski}}{{Udalski} et~al.}{2015}]{Udalski_etal_2015}
{Udalski} A.,  {Szyma{\'n}ski} M.~K.,   {Szyma{\'n}ski} G.,  2015, \actaa,
  \href {http://cdsads.u-strasbg.fr/abs/2015AcA....65....1U} {65, 1}

\bibitem[\protect\citeauthoryear{{Vogt}}{{Vogt}}{1990}]{1990ApJ...356..609V}
{Vogt} N.,  1990, \mn@doi [\apj] {10.1086/168866}, \href
  {https://ui.adsabs.harvard.edu/abs/1990ApJ...356..609V} {356, 609}

\bibitem[\protect\citeauthoryear{{Walter}}{{Walter}}{2016}]{Walter_2016}
{Walter} F.~M.,  2016, in Conference on Shocks and Particle Acceleration in
  Novae and Supernovae.
  \url{https://www.simonsfoundation.org/event/conference-on-shocks-and-particle-acceleration-in-novae-and-supernovae/}

\bibitem[\protect\citeauthoryear{{Walter}, {Battisti}, {Towers}, {Bond}  \&
  {Stringfellow}}{{Walter} et~al.}{2012}]{Walter_etal_2012}
{Walter} F.~M.,  {Battisti} A.,  {Towers} S.~E.,  {Bond} H.~E.,
  {Stringfellow} G.~S.,  2012, \mn@doi [\pasp] {10.1086/668404}, \href
  {http://cdsads.u-strasbg.fr/abs/2012PASP..124.1057W} {124, 1057}

\bibitem[\protect\citeauthoryear{{Wang} \& {Chen}}{{Wang} \&
  {Chen}}{2019}]{Wang_etal_2019}
{Wang} S.,  {Chen} X.,  2019, \mn@doi [\apj] {10.3847/1538-4357/ab1c61}, \href
  {https://ui.adsabs.harvard.edu/abs/2019ApJ...877..116W} {877, 116}

\bibitem[\protect\citeauthoryear{{Warner}}{{Warner}}{1987}]{Warner_etal_1987MNRAS}
{Warner} B.,  1987, \mn@doi [\mnras] {10.1093/mnras/227.1.23}, \href
  {https://ui.adsabs.harvard.edu/abs/1987MNRAS.227...23W} {227, 23}

\bibitem[\protect\citeauthoryear{{Warner}}{{Warner}}{1995}]{Warner_1995}
{Warner} B.,  1995, {Cataclysmic variable stars}.
 Vol. 28

\bibitem[\protect\citeauthoryear{{Warner}}{{Warner}}{2008}]{Warner_2008}
{Warner} B.,  2008, in {Bode} M.~F.,  {Evans} A.,  eds, Classical Novae. pp
  16--33

\bibitem[\protect\citeauthoryear{{Williams}}{{Williams}}{1992}]{Williams_1992}
{Williams} R.~E.,  1992, \mn@doi [\aj] {10.1086/116268}, \href
  {http://adsabs.harvard.edu/abs/1992AJ....104..725W} {104, 725}

\bibitem[\protect\citeauthoryear{{Williams}}{{Williams}}{2012}]{Williams_2012}
{Williams} R.~E.,  2012, \mn@doi [\aj] {10.1088/0004-6256/144/4/98}, \href
  {http://adsabs.harvard.edu/abs/2012AJ....144...98W} {144, 98}

\bibitem[\protect\citeauthoryear{{Williams}, {Darnley}, {Bode}  \&
  {Steele}}{{Williams} et~al.}{2015}]{2015ApJ...805L..18W}
{Williams} S.~C.,  {Darnley} M.~J.,  {Bode} M.~F.,   {Steele} I.~A.,  2015,
  \mn@doi [\apjl] {10.1088/2041-8205/805/2/L18}, \href
  {https://ui.adsabs.harvard.edu/abs/2015ApJ...805L..18W} {805, L18}

\bibitem[\protect\citeauthoryear{{Wolf}, {Bildsten}, {Brooks}  \&
  {Paxton}}{{Wolf} et~al.}{2013}]{Wolf_etal_2013}
{Wolf} W.~M.,  {Bildsten} L.,  {Brooks} J.,   {Paxton} B.,  2013, \mn@doi
  [\apj] {10.1088/0004-637X/777/2/136}, \href
  {http://cdsads.u-strasbg.fr/abs/2013ApJ...777..136W} {777, 136}

\bibitem[\protect\citeauthoryear{{Yaron}, {Prialnik}, {Shara}  \&
  {Kovetz}}{{Yaron} et~al.}{2005}]{Yaron_etal_2005}
{Yaron} O.,  {Prialnik} D.,  {Shara} M.~M.,   {Kovetz} A.,  2005, \mn@doi
  [\apj] {10.1086/428435}, \href
  {http://cdsads.u-strasbg.fr/abs/2005ApJ...623..398Y} {623, 398}

\bibitem[\protect\citeauthoryear{{Zemko}, {Mukai}  \& {Orio}}{{Zemko}
  et~al.}{2015}]{Zemko_etal_2015}
{Zemko} P.,  {Mukai} K.,   {Orio} M.,  2015, \mn@doi [\apj]
  {10.1088/0004-637X/807/1/61}, \href
  {http://cdsads.u-strasbg.fr/abs/2015ApJ...807...61Z} {807, 61}

\bibitem[\protect\citeauthoryear{{Zemko}, {Orio}, {Mukai}, {Bianchini}, {Ciroi}
   \& {Cracco}}{{Zemko} et~al.}{2016}]{Zemko_etal_2016}
{Zemko} P.,  {Orio} M.,  {Mukai} K.,  {Bianchini} A.,  {Ciroi} S.,   {Cracco}
  V.,  2016, \mn@doi [\mnras] {10.1093/mnras/stw1199}, \href
  {http://cdsads.u-strasbg.fr/abs/2016MNRAS.460.2744Z} {460, 2744}

\bibitem[\protect\citeauthoryear{{Zwitter} \& {Munari}}{{Zwitter} \&
  {Munari}}{1995}]{Zwitter_Munari_1995}
{Zwitter} T.,  {Munari} U.,  1995, \aaps, \href
  {https://ui.adsabs.harvard.edu/abs/1995A&AS..114..575Z} {114, 575}

\bibitem[\protect\citeauthoryear{{van den Heuvel}, {Bhattacharya}, {Nomoto}  \&
  {Rappaport}}{{van den Heuvel} et~al.}{1992}]{vandenHeuvel_etal_1992}
{van den Heuvel} E.~P.~J.,  {Bhattacharya} D.,  {Nomoto} K.,   {Rappaport}
  S.~A.,  1992, \aap, \href
  {https://ui.adsabs.harvard.edu/abs/1992A&A...262...97V} {262, 97}

\makeatother
\end{thebibliography}

\appendix

\renewcommand\thetable{\thesection.\arabic{table}}    
\renewcommand\thefigure{\thesection.\arabic{figure}}   
\setcounter{figure}{0}

\section{Supplementary plots and tables}
\label{appB}
In this Appendix we present supplementary plots and tables.

\begin{figure*}
\begin{center}
  \includegraphics[width=0.49\textwidth]{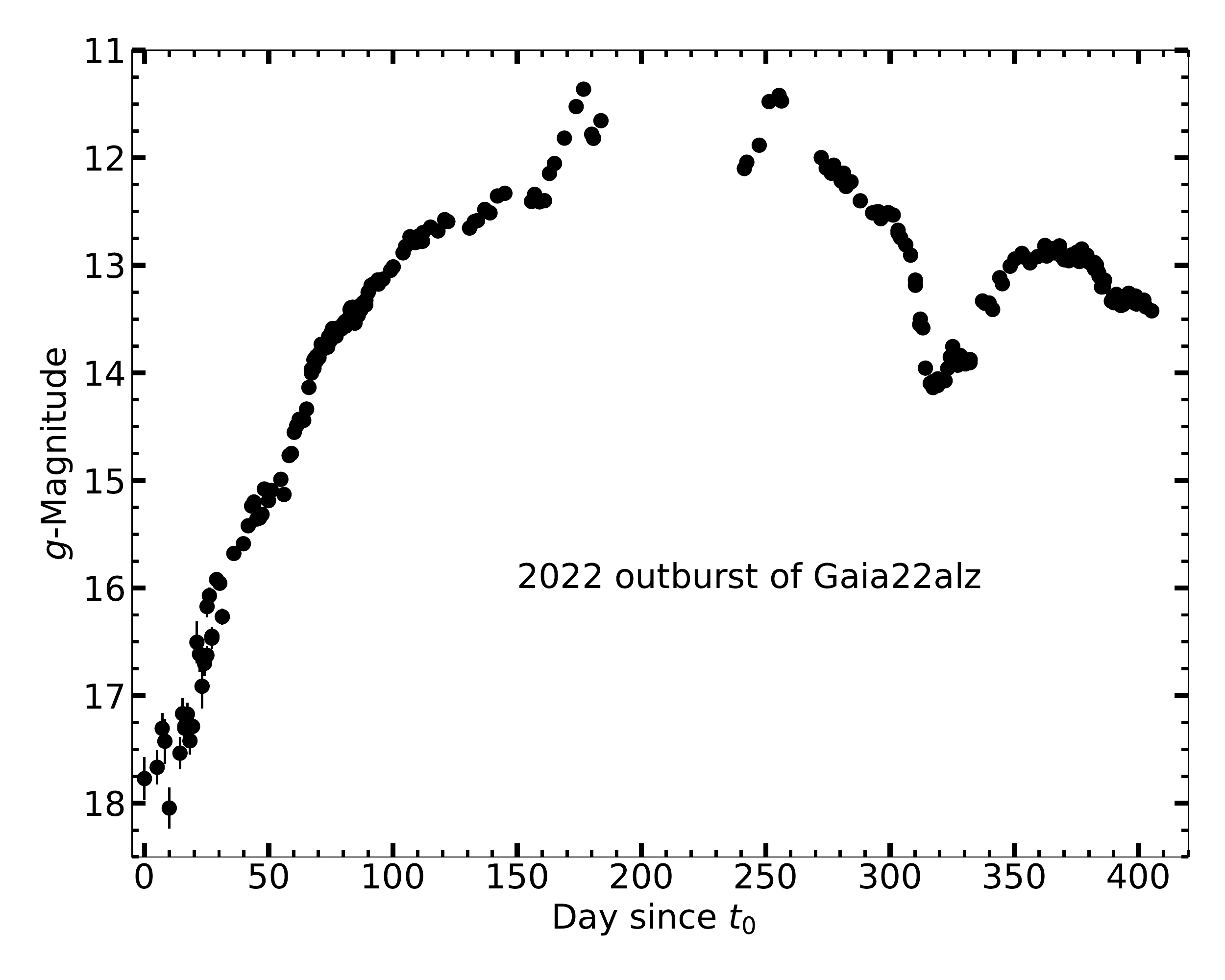}
    \includegraphics[width=0.49\textwidth]{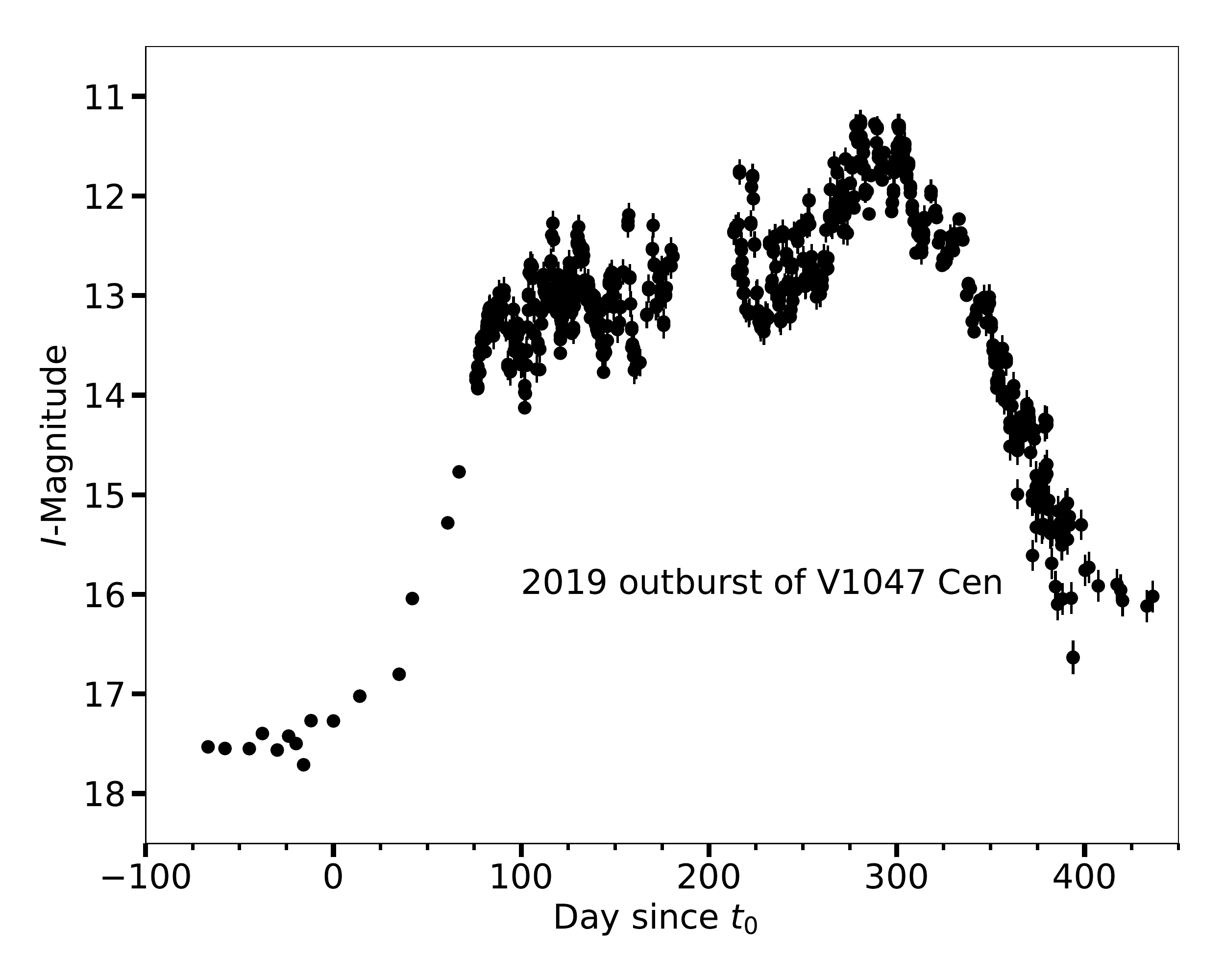}
\caption{\textit{Left}: the complete ASAS-SN optical $g$-band light curve of nova Gaia22alz, up until the time of the writing of this paper. \textit{Right}: the $I$-band light curve of the 2019 outburst of V1047~Cen, adopted from \citet{Aydi_etal_2022}.}
\label{Fig:LC_comp}
\end{center}
\end{figure*}

\begin{figure*}
\begin{center}
  \includegraphics[width=\textwidth]{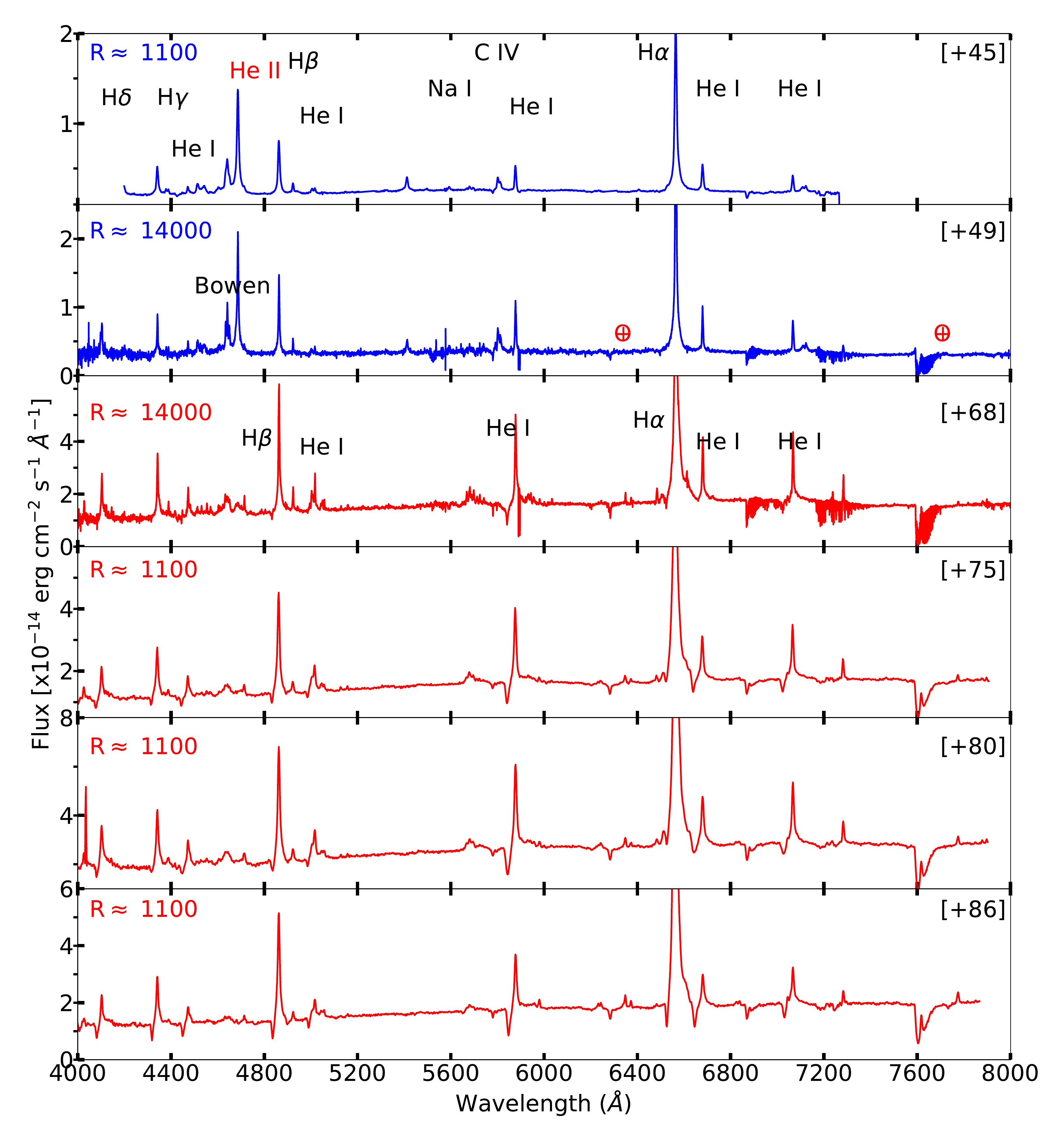}
\caption{The spectral evolution of nova Gaia22alz representing the different spectral stages: early-rise (blue), mid-rise (red), late-rise (green), and post-optical-peak (black). Numbers between brackets are days after $t_0$. The resolving power $R$ of each spectrum is added on the left-hand side of each panel.}
\label{Fig:spec_evolution_1}
\end{center}
\end{figure*}

\begin{figure*}
\begin{center}
  \includegraphics[width=\textwidth]{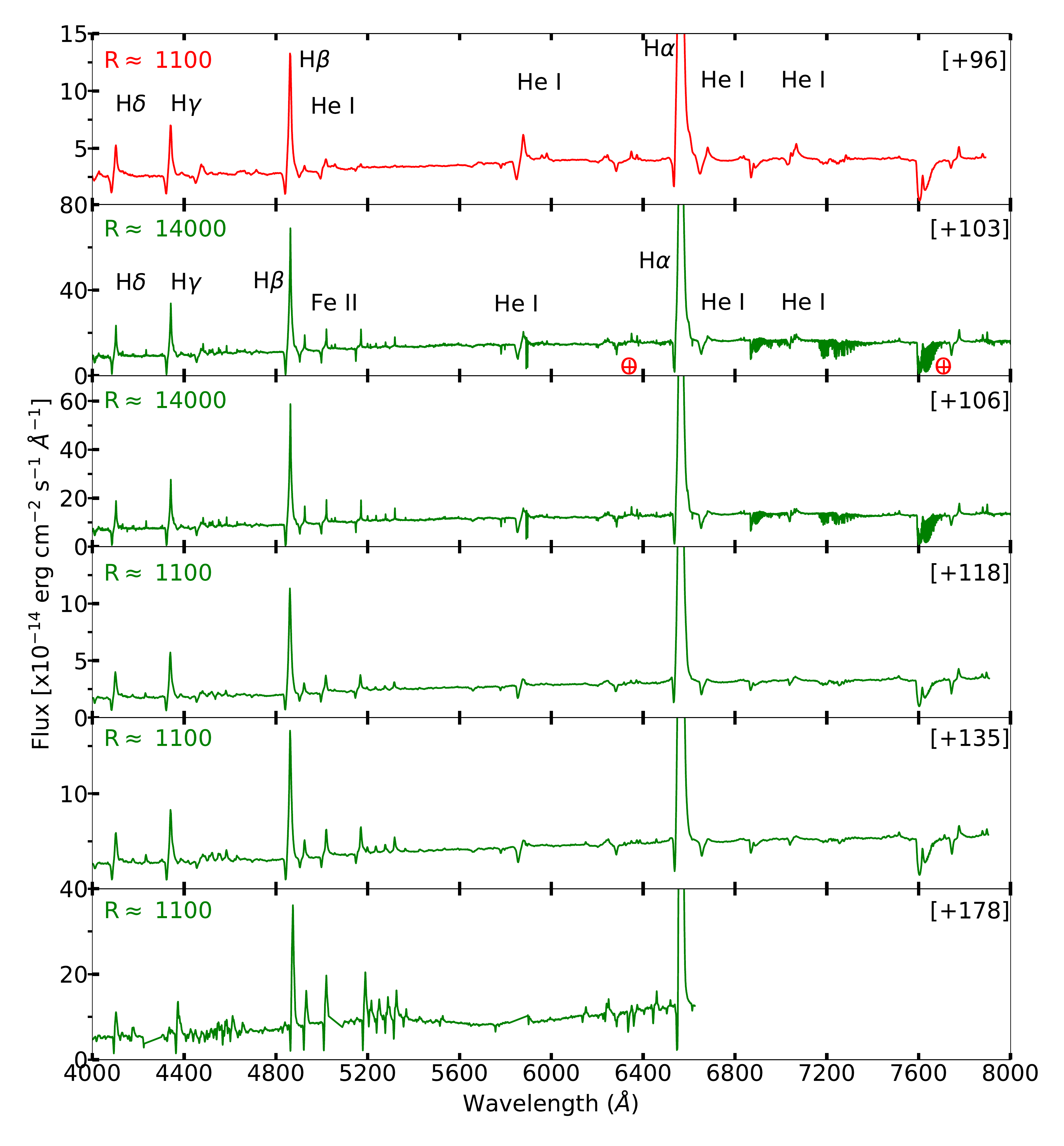}
\caption{The spectral evolution of nova Gaia22alz representing the different spectral stages: early-rise (blue), mid-rise (red), late-rise (green), and post-optical-peak (black). Numbers between brackets are days after $t_0$. The resolving power $R$ of each spectrum is added on the left-hand side of each panel.}
\label{Fig:spec_evolution_2}
\end{center}
\end{figure*}

\begin{figure*}
\begin{center}
  \includegraphics[width=\textwidth]{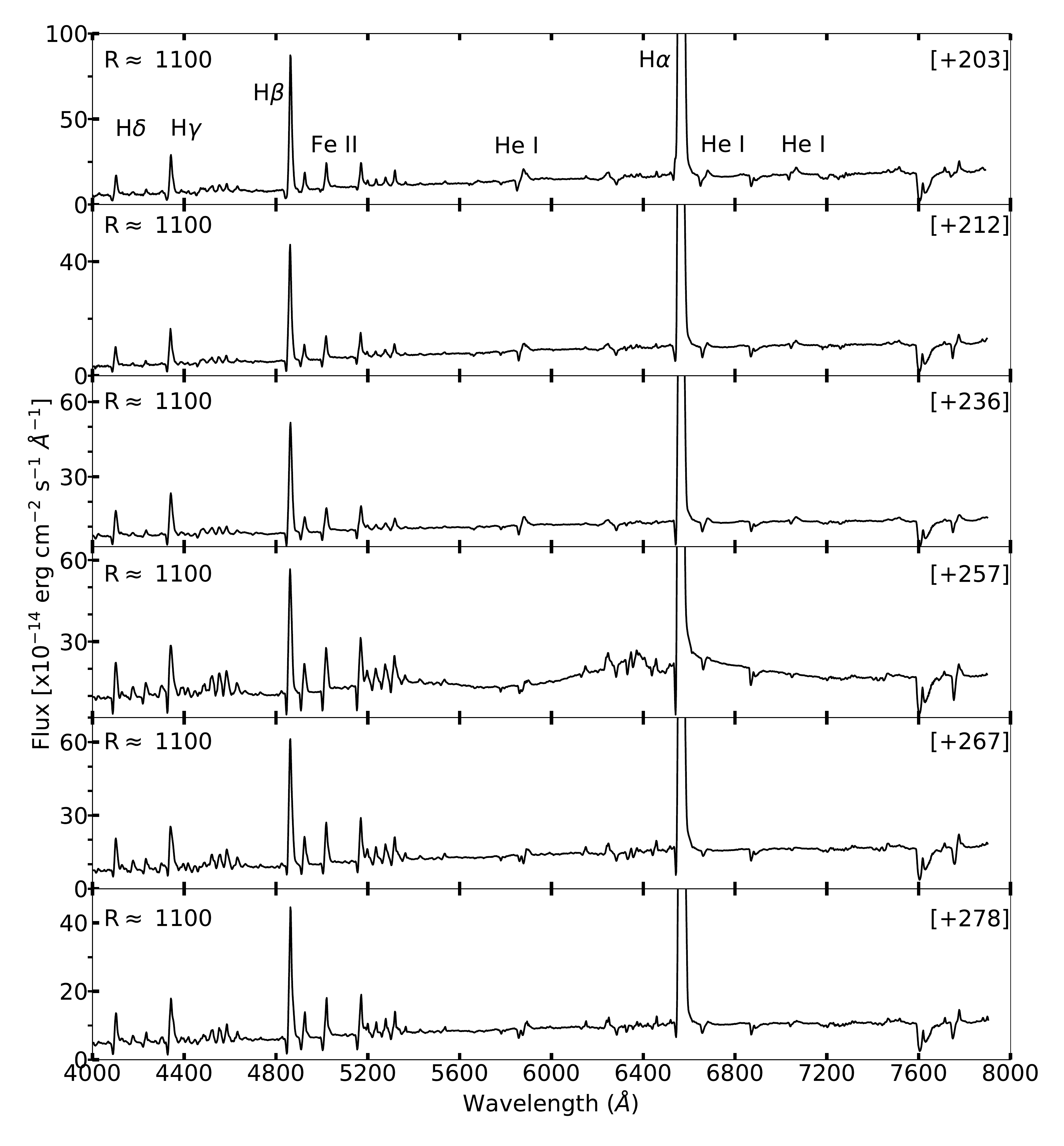}
\caption{The spectral evolution of nova Gaia22alz representing the different spectral stages: early-rise (blue), mid-rise (red), late-rise (green), and post-optical-peak (black). Numbers between brackets are days after $t_0$. The resolving power $R$ of each spectrum is added on the left-hand side of each panel.}
\label{Fig:spec_evolution_3}
\end{center}
\end{figure*}



\begin{table*}
\centering
\caption{Optical spectroscopic observations log.}
\begin{tabular}{cccccc}
\hline
 Telescope & Instrument & date & $ t - t_0$ & Resolving power & $\lambda$ Range \\ 
&  & &(days) &  & ($\mathrm{\AA}$)\\
\hline
SALT & RSS & 2022-03-10 & 45 & 1500 & 4200\,--\,7300\\
SALT & HRS & 2022-03-15 & 49 & 14000 & 4000\,--\,9000\\
SALT & HRS & 2022-04-03 & 68 & 14000 & 4000\,--\,9000\\
SOAR & Goodman & 2022-04-09 & 76 & 1100 & 3800\,--\,7800\\
SOAR & Goodman & 2022-04-10 & 76 & 4400 & 4500\,--\,5170\\
SOAR & Goodman & 2022-04-14 & 80 & 1100 & 3800\,--\,7800\\
SOAR & Goodman & 2022-04-21 & 86 & 1100 & 3800\,--\,7800\\
SOAR & Goodman & 2022-04-21 & 86 & 4400 & 4500\,--\,5170\\
SOAR & Goodman & 2022-05-01 & 96 & 1100 & 3800\,--\,7800\\
SOAR & Goodman & 2022-05-01 & 96 & 4400 & 4500\,--\,5170\\
SALT & HRS & 2022-05-08 & 103 & 14000 & 4000\,--\,9000\\
SALT & HRS & 2022-05-11 & 106 & 14000 & 4000\,--\,9000\\
SOAR & Goodman & 2022-05-22 & 118 & 1100 & 3800\,--\,7800\\
SOAR & Goodman & 2022-05-22 & 118 & 4400 & 4500\,--\,5170\\
SOAR & Goodman & 2022-05-24 & 120 & 1100 & 3800\,--\,7800\\
SOAR & Goodman & 2022-05-24 & 120 & 4400 & 4500\,--\,5170\\
SOAR & Goodman & 2022-06-09 & 135 & 1100 & 3800\,--\,7800\\
SOAR & Goodman & 2022-06-09 & 135 & 4400 & 4500\,--\,5170\\
Magellan & IMACS & 2022-07-21 & 178 & 1500 & 3800\,--\,6600\\
SOAR & Goodman & 2022-08-15 & 203 & 1100 & 3800\,--\,7800\\
SOAR & Goodman & 2022-08-15 & 203 & 4400 & 4500\,--\,5170\\
SOAR & Goodman & 2022-08-24 & 212 & 1100 & 3800\,--\,7800\\
SOAR & Goodman & 2022-08-24 & 212 & 4400 & 4500\,--\,5170\\
SOAR & Goodman & 2022-09-17 & 236 & 1100 & 3800\,--\,7800\\
SOAR & Goodman & 2022-09-17 & 236 & 4400 & 4500\,--\,5170\\
SOAR & Goodman & 2022-10-08 & 257 & 1100 & 3800\,--\,7800\\
SOAR & Goodman & 2022-10-08 & 257 & 4400 & 4500\,--\,5170\\
SOAR & Goodman & 2022-10-09 & 258 & 4400 & 4500\,--\,5170\\
SOAR & Goodman & 2022-10-18 & 267 & 1100 & 3800\,--\,7800\\
SOAR & Goodman & 2022-10-18 & 267 & 4400 & 4500\,--\,5170\\
SOAR & Goodman & 2022-10-29 & 278 & 1100 & 3800\,--\,7800\\
SOAR & Goodman & 2022-10-29 & 278 & 4400 & 4500\,--\,5170\\
SOAR & Goodman & 2022-12-01 & 310 & 1100 & 3800\,--\,7800\\
SOAR & Goodman & 2022-12-01 & 310 & 4400 & 4500\,--\,5170\\
SOAR & Goodman & 2022-12-22 & 331 & 1100 & 3800\,--\,7800\\
SOAR & Goodman & 2022-12-22 & 331 & 4400 & 4500\,--\,5170\\
\hline
\end{tabular}
\label{table:spec_log}
\end{table*}
\clearpage

\onecolumn
\section{Analytical model of stationary shock within the L2 outflow}
\label{appC}
In this Appendix we present a set of calculations supporting the main, alternative model with suggest in Section~\ref{Disc}.
A binary of period $P$ and mass $M$ has a semi-major axis given by
\begin{equation}
a \approx 3.5\times 10^{10}{\rm cm}\left(\frac{M}{M_{\odot}}\right)^{1/3}\left(\frac{P}{\rm hr}\right)^{2/3},
\end{equation}
with binary escape speed
\begin{equation}
v_{\rm esc} \approx \left(\frac{2GM}{a}\right)^{1/2} \approx 870\,{\rm km\,s^{-1}}\,\left(\frac{M}{M_{\odot}}\right)^{1/3}\left(\frac{P}{\rm hr}\right)^{-1/3}.
\end{equation}

Consider a mass-loss from the $L_2$ point at a rate $\dot{M}_{\rm L}$, which we assume is monotonically rising after the first few days of the eruption. \citet{Shu_etal_1979} and \citet{Pejcha_etal_2016} show that cold mass-loss from the $L_2$ point for a binary of $0.064 \lesssim q \equiv M_{\star}/M_{\rm WD} \lesssim 0.78$ reaches a terminal velocity up to
\begin{equation}
v_{\rm L2,min} \approx v_{\rm esc}/4 \approx 220\,{\rm km\,s^{-1}}\,\left(\frac{M}{M_{\odot}}\right)^{1/3}\left(\frac{P}{\rm hr}\right)^{-1/3}.
\end{equation}
Higher velocities are predicted for matter ejected with finite thermal temperature \citep{Pejcha_etal_2017}, so we take $v_{\rm L2} \approx 2 v_{\rm L2,min}$, 
to match the 400 km s$^{-1}$ line widths observed in Gaia22alz.  The vertical surface density and optical depth of the equatorial outflow at radius $r \gg a$ from the binary is given by
\begin{equation}
\Sigma \approx \frac{\dot{M}_{\rm L2}}{2\pi r v_{\rm L2}}; \,\,\, \tau_{\rm T} \approx \frac{\dot{M}_{\rm L2}\kappa_{\rm T}}{2\pi r v_{\rm L2}} \approx 1.0\left(\frac{\dot{M}_{\rm L2}}{10^{-7}M_{\odot}{\rm wk}^{-1}}\right)\left(\frac{r}{10a}\right)^{-1}\left(\frac{M}{M_{\odot}}\right)^{-2/3}\left(\frac{P}{\rm hr}\right)^{-1/3}
\label{eq:tau}
\end{equation}
where $\kappa_{\rm T} \approx 0.34$ cm$^{2}$ g$^{-1}$ is the Thomson optical depth. The $L_2$ streams collide at a radius $r_{\rm coll} \approx 10a_{\rm bin}$, with a relative velocity $\Delta v \approx 0.2 v_{\rm L2}$, shock-heating the material to a temperature
\begin{equation}
T_{\rm c} \simeq \frac{3}{16}\frac{m_p}{k} (\Delta v)^{2} \approx 1.7\times 10^{5}\,{\rm K}\left(\frac{M}{M_{\odot}}\right)^{2/3}\left(\frac{P}{\rm hr}\right)^{-2/3},
\end{equation}
high enough to (re-)ionize the gas. The midplane density on the radial scale of the collision can likewise be estimated as \citet{Pejcha_etal_2016}
\begin{equation}
\rho_{\rm c} \sim \frac{\Sigma}{2a_{\rm bin}} \sim 5\times 10^{-11}{\rm g\,cm^{-3}}\left(\frac{\dot{M}_{\rm L2}}{10^{-7}M_{\odot}{\rm wk}^{-1}}\right)\left(\frac{r_{\rm coll}}{10a}\right)^{-1}\left(\frac{M}{M_{\odot}}\right)^{-1}\left(\frac{P}{\rm hr}\right)^{-1}.
\end{equation}
For typical values of $(\rho_{\rm c}, T_{\rm c}$), we estimate the opacity (e.g., from OPAL) to be close to the Thomson value assumed above (e.g., figure 3 of \citealt{Matsumoto_Metzger_2022}; see also below). The kinetic luminosity of the shocks is given by
\begin{equation}
L_{L2} \approx \dot{M}_{\rm L2}v_{\rm L2}\Delta v \approx 1.3\times 10^{35}\,{\rm erg\,s^{-1}}\left(\frac{\dot{M}_{\rm L2}}{10^{-7}M_{\odot}{\rm wk}^{-1}}\right)\left(\frac{M}{M_{\odot}}\right)^{2/3}\left(\frac{P}{\rm hr}\right)^{-2/3},
\label{eq:L2}
\end{equation}  
while the radiative cooling time of the gas,
\begin{equation}
t_{\rm cool} \sim \frac{3kT_{\rm c}}{(4\rho_{\rm c}/m_p)\Lambda(T_{\rm c})} \sim 10^{-3}\,{\rm s}\left(\frac{\dot{M}_{\rm L2}}{10^{-7}M_{\odot}{\rm wk}^{-1}}\right)^{-1},
\end{equation}
is extremely short, where $\Lambda(T) \simeq 2.8\times 10^{-18}(T/K)^{-0.7}$ erg s$^{-1}$ cm$^{3}$ is the atomic line cooling rate (e.g., \citealt{Draine_2009}).

Using Eq.~\ref{eq:tau}, \ref{eq:L2} we can eliminate $\dot{M}_{\rm L2}$ to write
\begin{equation}
L_{L2} \approx 10^{35}{\rm erg\,s^{-1}}\left(\frac{\tau_{\rm T}}{1}\right)\left(\frac{M}{M_{\odot}}\right)^{4/3}\left(\frac{P}{\rm hr}\right)^{-1/3},
\end{equation}
Up to what value of $\tau_{\rm T}$ can we still see line emission from the shock-heated gas?  Considering that the free-free opacity for an optical photon (e.g., H II) of energy $\epsilon_{\gamma} \sim 1$ eV,
\begin{equation}
\kappa_{\rm ff} \approx 0.35\,{\rm cm^{2}\,g^{-1}}\left(\frac{T}{10^{5}{\rm K}}\right)^{-3/2}\left(\frac{\rho_{\rm c}}{10^{-10}{\rm g\,cm^{-3}}}\right)\left(\frac{\epsilon_{\gamma}}{1\,{\rm eV}}\right)^{-2}\bar{g}_{\rm ff},
\end{equation}
is close to the Thomson opacity. Therefore, the effective absorption optical depth
$
\tau_{\rm th} \simeq \sqrt{\tau_{\rm ff}(\tau_{\rm T}+\tau_{\rm ff})}
$
is close to $\tau_{\rm T}$.  In other words, we expect to observe line emission up to $\tau_{\rm T} \sim 1$, or $L_{\rm L2} \sim 10^{35}$ erg s$^{-1}$.

\end{document}